\documentclass{article}
\PassOptionsToPackage{hyphens}{url}\usepackage{hyperref}
\usepackage{graphicx}
\usepackage{amssymb}
\usepackage{amsmath}
\usepackage{amsfonts}
\usepackage{parskip}
\usepackage{titling}
\usepackage{color}
\usepackage{empheq}
\usepackage{xspace}

\newcommand{\rmh}[1]{#1}
\newcommand{\lb}[1]{#1}

\pretitle{\Large\bf} \posttitle{\par\vskip 0.3em} \preauthor{\large} \postauthor{\par\vskip 0.0em}
\predate{\large} \postdate{\par\vskip -1.1em \rule{\linewidth}{1pt}} 

\newcommand{\beq}{\begin{equation}}
\newcommand{\eeq}{\end{equation}}
\newcommand{\bea}{\begin{eqnarray}}
\newcommand{\eea}{\end{eqnarray}}
\newcommand{\norm}[1]{\lVert{#1}\rVert}
\newcommand{\comment}[1]{}
\renewcommand{\d}{{\rm d}}

\begin{document}

\title{Large-Scale Optical Neural Networks based on Photoelectric Multiplication}
\author{Ryan Hamerly, Liane Bernstein, Alexander Sludds, Marin Solja\v{c}i\'{c}, and Dirk Englund}

\date{May 16, 2019}
\maketitle

\begin{flushleft}
\textit{Research Laboratory of Electronics, MIT, 50 Vassar Street, Cambridge, MA 02139, USA}
\end{flushleft}

\begin{abstract}
Recent success in deep neural networks has generated strong interest in hardware accelerators to improve speed and energy consumption.  This paper presents a new type of photonic accelerator based on coherent detection that is scalable to large ($N \gtrsim 10^6$) networks and can be operated at high (GHz) speeds and very low (sub-aJ) energies per multiply-and-accumulate (MAC), using the massive spatial multiplexing enabled by standard free-space optical components.  In contrast to previous approaches, both weights and inputs are optically encoded so that the network can be reprogrammed and trained on the fly.  Simulations of the network using models for digit- and image-classification reveal a ``standard quantum limit'' for optical neural networks, set by photodetector shot noise.  This bound, which can be as low as 50~zJ/MAC, suggests performance below the thermodynamic (Landauer) limit for digital irreversible computation is theoretically possible in this device.  The proposed accelerator can implement both fully-connected and convolutional networks.  We also present a scheme for back-propagation and training that can be performed in the same hardware.  This architecture will enable a new class of ultra-low-energy processors for deep learning.
\end{abstract}

In recent years, deep neural networks have tackled a wide range of problems including image analysis \cite{Krizhevsky2012}, natural language processing \cite{Young2018}, game playing \cite{Silver2017}, physical chemistry \cite{Gilmer2017}, and medicine \cite{Wang2016}.  This is not a new field, however.  The theoretical tools underpinning deep learning have been around for several decades \cite{Rosenblatt1958, Werbos1974, LeCun1998}; the recent resurgence is driven primarily by (1) the availability of large training datasets \cite{ILSVRC15}, and (2) substantial growth in computing power \cite{Moore1965} and the ability to train networks on GPUs \cite{Steinkraus2005}.  Moving to more complex problems and higher network accuracies requires larger and deeper neural networks, which in turn require even more computing power \cite{Canziani2017}.  This motivates the development of special-purpose hardware optimized to perform neural-network inference and training \cite{Sze2017}.  

To outperform a GPU, a neural-network accelerator must significantly lower the energy consumption, since the performance of modern microprocessors is limited by on-chip power \cite{Horowitz2014}.  In addition, the system must be fast, programmable, scalable to many neurons, compact, and ideally compatible with training as well as inference.  Application-specific integrated circuits (ASICs) are one obvious candidate for this task.  State-of-the-art ASICs can reduce the energy per multiply-and-accumulate (MAC) from 20 pJ/MAC for modern GPUs \cite{Keckler2011} to around 1 pJ/MAC \cite{Chen2014, Jouppi2017}.  However, ASICs are based on CMOS technology and therefore suffer from the interconnect problem---even in highly optimized architectures where data is stored in register files close to the logic units, a majority of the energy consumption comes from data movement, not logic \cite{Sze2017, Chen2014}.  Analog crossbar arrays based on CMOS gates \cite{George2016} or memristors \cite{Kim2011, Li2018} promise better performance, but as analog electronic devices, they suffer from calibration issues and limited accuracy \cite{Feinberg2018}.  


Photonic approaches can greatly reduce both the logic and data-movement energy by performing (the linear part of) each neural-network layer in a passive, linear optical circuit.  This allows the linear step is performed at high speed with no energy consumption beyond transmitter and receiver energies.  Optical neural networks based on free-space diffraction \cite{Lin2018} have been reported, but require spatial light modulators or 3D-printed diffractive elements, and are therefore not rapidly programmable.  Nanophotonic circuits are a promising alternative \cite{Shen2017, Tait2017}, but the footprint of directional couplers and phase modulators makes scaling to large ($N \geq 1000$) numbers of neurons very challenging.  To date, the goal of a large-scale, rapidly reprogrammable photonic neural network remains unrealized.

This paper presents a new architecture based on coherent (homodyne) detection that is fast, low-power, compact, and readily scalable to large ($N \gtrsim 10^6$) numbers of neurons.  In contrast to previous schemes, here we encode both the inputs and weights in optical signals, allowing the weights to be changed on the fly at high speed.  Synaptic connections (matrix-vector products) are realized by the {\it quantum photoelectric multiplication} process in the homodyne detectors.  Our system is naturally adapted to free-space optics and can therefore take advantage of the massive spatial multiplexing possible in free-space systems \cite{Kahn2017, Miller2017} and the high pixel density of modern focal-plane arrays \cite{Rogalski2012} to scale to far more neurons than can be supported in nanophotonics or electronic cross-bar arrays.  The optical energy consumption is subject to a fundamental standard quantum limit (SQL) arising from the effects of shot noise in photodetectors, which lead to classification errors.  Simulations based on MNIST neural networks empirically show the SQL can be as low as 50--100 zJ/MAC.  Using realistic laser, modulator, and detector energies, performance at the sub-fJ/MAC level should be possible with present technology.  The optical system can be used for both fully-connected and convolutional layers.  Finally, backpropagation is straightforward to implement in our system, allowing both inference and training to be performed in the same optical device.

\begin{figure}[b!]
\begin{center}
\includegraphics[width=1.00\textwidth]{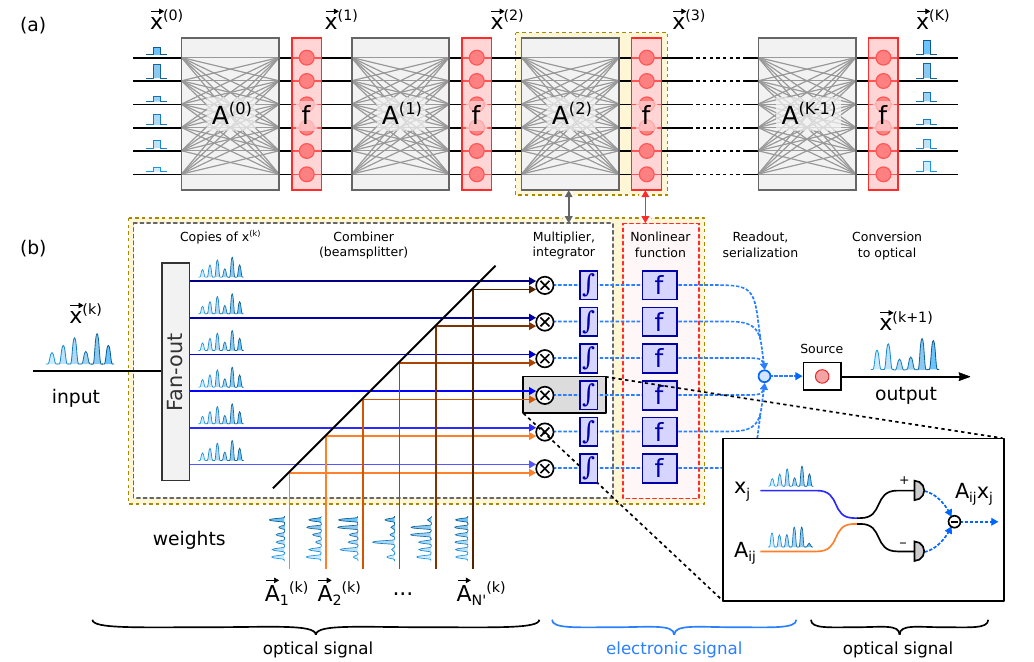}
\caption{Schematic diagram of a single layer of the homodyne optical neural network.  (a) Neural network represented as a sequence of $K$ layers, each consisting of a matrix-vector multiplication (grey) and an element-wise nonlinearity (red).  (b) Implementation of a single layer.  Matrix multiplication is performed by combining input and weight signals and performing \rmh{balanced} homodyne detection (inset) between each signal-weight pair (grey box).  For details on experimental implementation see Supp.~Sec.~\ref{sec:s1}.  The resulting electronic signals are sent through a nonlinear function (red box), serialized, and send to the input of the next layer.}
\label{fig:f1}
\end{center}
\end{figure}

\subsection*{Coherent Matrix Multiplier}

Fig.~\ref{fig:f1} illustrates the device.  A deep neural network is a sequence of $K$ layers (Fig.~\ref{fig:f1}(a)), each consisting of a matrix multiplication $\vec{x}\rightarrow A\vec{x}$ (synaptic connections) and an element-wise nonlinearity $x_i \rightarrow f(x_i)$ (activation function); thus the input into the $(k+1)^{\rm th}$ layer is related to the $k^{\rm th}$ layer input by:
\beq
	x_i^{(k+1)} = f\Bigl(\sum_j A_{ij}^{(k)} x_j^{(k)}\Bigr) \label{eq:nnlayer}  
\eeq
For a given layer, let $N$ and $N'$ be the number of input and output neurons, respectively.  Input (output) data are encoded temporally as $N$ ($N'$) pulses on a single channel as shown in Fig.~\ref{fig:f1}(b).  This encoding, reminiscent of the Coherent Ising Machine \cite{Marandi2014, Inagaki2016, McMahon2016}, contrasts with other approaches used for neural networks, which encode inputs in separate spatial channels \cite{Shen2017, Tait2017, Lin2018}.  As there are $NN'$ weights for an $N'\times N$ fully-connected matrix, the weights enter on $N'$ separate channels, each carrying a single matrix row encoded in time.  Input data is optically fanned out to all $N'$ channels, and each detector functions as a quantum photoelectric multiplier, calculating the homodyne product between the two signals (inset).  As long as both signals are driven from the same coherent source and the path-length difference is less than the coherence length, the charge $Q_i$ accumulated by homodyne receiver $i$ is:
\beq
	Q_i = \frac{2\eta e}{\hbar\omega} \int{\mbox{Re}\bigl[E^{\rm (in)}(t)^* E_i^{\rm (wt)}(t)\bigr]\d t}
		\propto \sum_j A_{ij} x_j
\eeq
Here $E^{\rm (in)}(t)$ and $E_i^{\rm (wt)}(t)$ are the input and weight fields for receiver $i$, which are taken to be sequences of pulses with amplitudes proportional to $x_j$ and $A_{ij}$, respectively ($x_j, A_{ij} \in \mathbb{R}$).  Thus each receiver performs a vector-vector product between $\vec{x}$ and a row $\vec{A}_i$ of the weight matrix; taken together, the $N'$ electronic outputs give the matrix-vector product $A\vec{x}$.  Fields are normalized so that power is given by $P(t) = |E(t)|^2$, and $\eta$ is the detector efficiency.  A serializer reads out these values one by one, applies the nonlinear function $f(\cdot)$ in the electrical domain, and outputs the result to a modulator to produce the next layer's inputs.

The \rmh{balanced} homodyne detector in Fig.~\ref{fig:f1}(b) (inset) combines the advantages of optics and electronics: it can process data encoded at extremely high speeds, limited only by the bandwidth of the beamsplitter ($\gtrsim$THz) and the (optical) bandwidth of the photodetectors (typically $\gtrsim$\,100 nm, or $\gtrsim$\,10 THz).  
The electrical bandwidth can be much slower, since only the integrated charge is measured.  Finally, the present scheme avoids the need for \rmh{low-power} nonlinear optics that is a major stumbling block in all-optical logic \cite{Miller2010b}: since the output is electrical, the dot product $A_{ij} x_j$ can be computed at extremely low power (sub-fJ/MAC) using standard non-resonant components (photodiodes) that are CMOS-compatible and scalable to arrays of millions.

\rmh{Previous approaches used optoelectronics (photodiodes, lasers, amplifiers) both to sum neuron inputs \cite{Tait2014, Tait2017} and to generate nonlinearity or spiking dynamics \cite{Vandoorne2008, Paquot2012, Larger2012, Nahmias2013, Brunner2016}; here, thanks to the optical weight encoding, the synaptic weighting itself is performed opto-electronically.}

Coherent detection greatly simplifies the setup compared to alternative approaches.  With a given set of weight inputs, the network in Fig.~\ref{fig:f1}(b) requires $N$ input pulses and $N'$ detectors to perform a matrix-vector operation with $NN'$ MACs, performing an operation that should scale quadratically with size using only linear resources.  This is in contrast to electrical approaches that require quadratic resources ($NN'$ floating-point operations total).  The (optical) energy consumption of nanophotonic systems \rmh{\cite{Shen2017, Tait2017}} also scales linearly for the same operation; however, the circuit is much more complex, requiring $O(NN')$ tunable phase shifters \cite{Reck1994, Clements2016} \rmh{or ring resonators \cite{Tait2017}}, which becomes very challenging to scale beyond several hundred channels and may be sensitive to propagation of fabrication errors.  The main caveat to our system is the need to generate the weights in the first place, which imposes an energy cost that does scale quadratically.  However, in many cases (particularly in data centers) neural networks are run simultaneously over large batches of data, so with appropriate optical fan-out, the cost of the weights can be amortized over many clients.  Put another way, running the neural network on data with batch size $B$, we are performing a matrix-matrix product $Y_{N'\times B} = A_{N'\times N} X_{N\times B}$, which requires $N'N B$ MACs, with an energy cost that should scale as $O(N'N) + O(N'B) + O(N B)$ rather than $O(N'N B)$.

\subsection*{Deep Learning at the Standard Quantum Limit}

As energy consumption is a primary concern in neuromorphic and computing hardware generally \cite{Horowitz2014}, an optical approach must outperform electronics by a large factor to justify the investment in a new technology.  In addition, optical systems must show great potential for improvement, ideally by many orders of magnitude, to allow continued scaling beyond the physical limits of Moore's Law.  Thus two questions are relevant: (1) the fundamental, physical limits to the energy consumption, and (2) the energy consumption of a practical, near-term device using existing technology.

The fundamental limit stems from quantum-limited noise.  In an electrical signal, energy is quantized at a level $E_{\rm el} = h/\tau_{\rm el}$, where $\tau_{\rm el} \sim 10^{-10}\,{\rm s}$ is the signal duration.  Optical energy is quantized at a level $E_{\rm opt} = h/\tau_{\rm opt}$, where $\tau_{\rm opt} \equiv c/\lambda \sim (2$--$5)\times 10^{-15}\,{\rm s}$, which is $10^4$--$10^5$ times higher.  As a result, $E_{\rm opt} \gg kT \gg E_{\rm el}$ and electrical signals can be treated in a classical limit governed by thermal noise, while optical signals operate in a zero-temperature quantum limit where vacuum fluctuations dominate.  These fluctuations are read out on the photodetectors, where the photoelectric effect \cite{Einstein1905} produces a Poisson-distributed photocurrent \cite{WallsMilburn, Hayat1996}.  While the photocurrents are subtracted in homodyne detection, the fluctuations add in quadrature, and Eq.~(\ref{eq:nnlayer}) is replaced by (See Supp.~Sec.~\ref{sec:s3} for derivation and assumptions):
\beq
	x_i^{(k+1)} = f\biggl(\sum_j A_{ij}^{(k)} x_j^{(k)} + w_i^{(k)} \frac{\norm{A^{(k)}}\norm{x^{(k)}}}{\sqrt{N^2 N'}} \frac{\sqrt{N}}{\sqrt{n_{\rm mac}}} \biggr) \label{eq:nnlayer2}  
\eeq
Here the $w_i^{(k)} \sim N(0, 1)$ are Gaussian random variables, $\norm{\cdot}$ is the $L_2$ norm, and $n_{\rm mac}$ is the number of photons per MAC, related to the total energy consumption of the layer by $n_{\rm tot} = NN' n_{\rm mac}$.

\begin{figure}[tb]
\begin{center}
\includegraphics[width=1.00\textwidth]{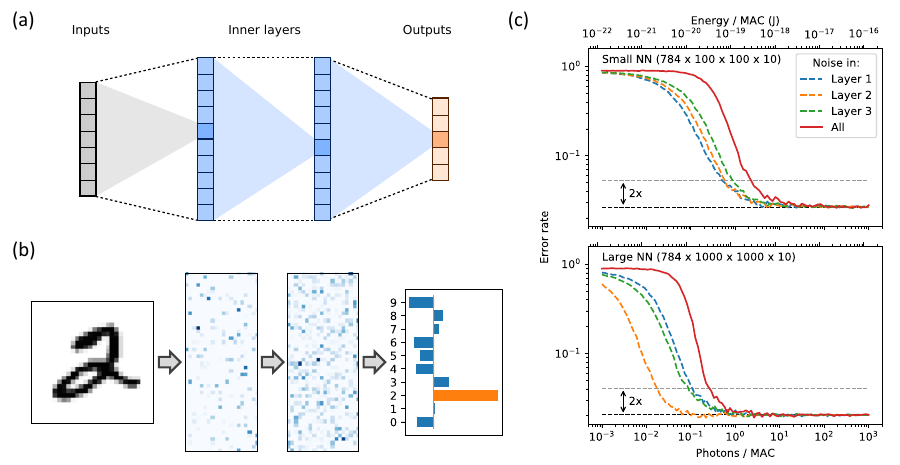}
\caption{(a) Illustration of a 3-layer neural network with full connectivity.  (b) MNIST image classified by network (size $784 \rightarrow 1000 \rightarrow 1000 \rightarrow 10$).  (c) Error rate as a function of photons per MAC $n_{\rm mac}$ (equivalently energy $E_{\rm mac} = (hc/\lambda) n_{\rm mac}$, here $\lambda = 1.55\mu$m).}
\label{fig:f2}
\end{center}
\end{figure}

The noise term in Eq.~(\ref{eq:nnlayer2}) scales as $n_{\rm mac}^{-1/2}$, and therefore the signal-to-noise ratio (SNR)  of each layer will scale as ${\rm SNR} \propto n_{\rm mac}$.  Since noise adversely affects the network's performance, one expects that the energy minimum should correspond to the value of $n_{\rm mac}$ at which the noise becomes significant.  To quantify this statement, we perform benchmark simulations using a collection of neural networks trained on the MNIST (digit recognition) dataset.  While MNIST digit classification is a relatively easy task \cite{Sze2017}, the intuition developed here should generalize to more challenging problems.  Data for two simple networks are shown in Fig.~\ref{fig:f2}, both having a 3-layer, fully-connected topology (Fig.~\ref{fig:f2}(a)).  In the absence of noise, the networks classify images with high accuracy, as the example illustrates (Fig.~\ref{fig:f2}(b)).

As Fig.~\ref{fig:f2}(c) shows, the error rate is a monotonically decreasing function of $n_{\rm mac}$.  The two asymptotic limits correspond to the noiseless case ($n_{\rm mac} \rightarrow \infty$, which returns the network's canonical accuracy), and the noise-dominated case ($n_{\rm mac} \rightarrow 0$, where the network is making a random guess).  Of interest to us is the cutoff point, loosely defined as the lowest possible energy at which the network returns close to its canonical accuracy \rmh{(for example within a factor of $2\times$, see dashed lines in Fig.~\ref{fig:f2}(c))}.  This is around 0.5--1 aJ (5--10 photons) for the small network (inner layer size $N = 100$), and 50--100 zJ (0.5--1 photon) for the large network (inner layer size $N = 1000$).  \rmh{(Note that this is per MAC, the number of photons per detector $N n_{\rm mac}$ is typically $\gg 1$.)}  This bound stems from the standard quantum limit (SQL): the intrinsic uncertainty of quadrature measurements on coherent states \cite{Caves1981}, which is temperature- and device-independent.  This should be viewed as an absolute lower bound for the energy consumption of neural networks of this type; although the use of squeezed light allows one to reach sensitivity below the SQL \cite{Jaekel1990, Grote2013}, this requires squeezing all inputs (including vacuum inputs \rmh{in optical fan-out}) which will likely lead to a net increase in overall energy consumption \rmh{(squeezing injects an average of $\sinh^2(\eta)$ photons per pulse, where $\eta$ is the squeezing parameter \cite{WallsMilburn}, which will substantially increase $n_{\rm mac}$)}.

The SQL is network-dependent, and not all layers contribute equally.  For each MAC, we have ${\rm SNR} \propto n_{\rm mac}$; however, the signal adds linearly while the errors add in quadrature.  As a result, the larger network is more resilient to individual errors because each output is averaging over more neurons.  Moreover, the solid curves in Fig.~\ref{fig:f2}(c) are restricted to the case when $n_{\rm mac}$ is the same for all layers.  The dashed lines show the error rate in a fictitious device where quantum-limited noise is only present in a particular layer.  For the large network, a smaller $n_{\rm mac}$ can be tolerated in the second layer, suggesting that better performance could be achieved by \rmh{independently} tuning the energy for each layer.  \rmh{Moreover, just as neural networks can be ``co-designed'' to achieve high accuracy on limited bit-precision hardware \cite{Sze2017}, changes to the training procedure (e.g.\ injecting noise to inner layers, a technique used to reduce generalization error \cite{Holmstrom1992, Hinton2012}) may further improve performance at low powers.}

Quantum limits to computational energy efficiency in photonics are not unique to neural networks.  In digital photonic circuits based on optical bistability \cite{Gibbs2012}, vacuum fluctuations lead to spontaneous switching events that limit memory lifetime and gate accuracy \cite{Savage1988, Santori2014}.  However, these effects require bistability at the attojoule scale \cite{Savage1988, Kerckhoff2011}, which is well out of the reach of integrated photonics (although recent developments are promising \cite{Ji2017, Hu2018, Wang2018}).  By contrast, neural networks are analog systems so the quantum fluctuations set a meaningful limit on efficiency even though no attojoule-scale optical nonlinearities are employed.

\subsection*{Energy Budget}

\begin{figure}[tb]
\begin{center}
\includegraphics[width=1.00\textwidth]{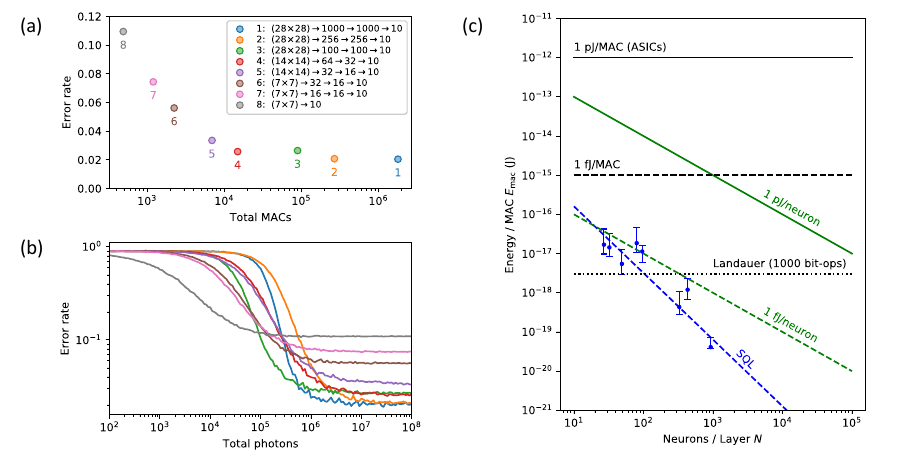}
\caption{(a) Conventional picture: Error rate as a function of number of MACs for different fully-connected MNIST neural networks.  (b) SQL picture: error rate as a function of total number of photons, for the same networks.  (c) Contributions to energy budget.  SQL dots correspond to minimum $E_{\rm mac}$ required to make the error rate $p_{\rm err}(E_{\rm mac}) < 1.5 p_{\rm err}(\infty)$ (error bars correspond to $p_{\rm err}(E_{\rm mac}) = [1.2, 2.0] p_{\rm err}(\infty)$).  $E_{\rm mac} = n_{\rm mac}(hc/\lambda)$, $\lambda = 1.55\mu$m.}
\label{fig:f3}
\end{center}
\end{figure}

Viewing the neural network as an analog system with quantum-limited performance shifts the paradigm for comparing neural networks.  Fig.~\ref{fig:f3}(a) shows the standard approach: a scatterplot comparing error rate with number of MACs, a rough proxy for time or energy consumption \cite{Canziani2017, Sze2017}.  There is a tradeoff between size and accuracy, with larger networks requiring more operations but also giving better accuracy.  In the SQL picture, each point becomes a curve because now we are free to vary the number of photons per MAC, and the energy bound is set by the total number of photons, not the number of MACs.  Fig.~\ref{fig:f3}(b) plots the error rate as a function of photon number for the networks above.  While the general tradeoff between energy and accuracy is preserved, there are a number of counterintuitive results.  For example, according to Fig.~\ref{fig:f3}(a), networks 1 and 2 have similar performance but the first requires $8\times$ more MACs, so under a conventional analysis, network 2 would always be preferred.  However, Fig.~\ref{fig:f3}(b) indicates that network 1 has better performance at all energy levels.  This is because network 1 is less sensitive to shot noise due to averaging over many neurons, and therefore can be operated at lower energies, compensating for the increased neuron count.  The same apparent paradox is seen with networks 3 and 4.  This suggests that, in a quantum-limited scenario, reducing total energy may not be as simple as reducing the number of operations.

The total energy budget depends on many factors besides the SQL. Fig.~\ref{fig:f3}(c) plots energy per MAC as a function of the average number of input neurons per layer $N$, a rough ``size'' of the neural network.  The SQL data are plotted for the eight networks in Fig.~\ref{fig:f3}(a-b), and the corresponding dashed line is an empirical fit.  Note that the SQL is an absolute lower bound, assumes perfect detectors, and only counts input optical energy.  In a realistic device, this curve is shifted up by a factor $(\eta_d \eta_c \eta_s \beta_{\rm mod})^{-1}$, where $\eta_d$, $\eta_c$, and $\eta_s$ are the detector, coupling, and source (laser) efficiencies and $\beta_{\rm mod}$ is the modulator launch efficiency \cite{Miller2012}; these are all close enough to unity in integrated systems \cite{Miller2017, Sun2015, Atabaki2018, Michaels2018} that the factor is $\lesssim 10$.

Another key factor is the detector electronics.  The homodyne signal from each neuron needs to be sent through a nonlinear function $y_i \rightarrow f(y_i)$ and converted to the optical domain using a modulator (Fig.~\ref{fig:f1}(b)).  The most obvious way to do this is to \rmh{amplify and} digitize the signal, perform the function $f(\cdot)$ in digital logic, serialize the outputs, convert back to analog, and send the analog signal into the modulator.  \rmh{Transimpedance amplifiers designed for optical interconnects operate at the $\sim100$ fJ range \cite{Saeedi2016, Miller2017}, while} ADCs in the few-pJ/sample regime are available \cite{Jonsson2011} and simple arithmetic \rmh{(for the activation function)} can be performed at the pJ scale \cite{Keckler2011, Chen2014, Jouppi2017}.  Modulators in this energy range are standard \cite{Sun2015, Atabaki2018, Saeedi2016}.  Thus a reasonable near-term estimate would be few-pJ/neuron; this figure is divided by the number of inputs per neuron to give the energy per MAC (solid green curve in Fig.~\ref{fig:f3}(c)).  \rmh{This few-pJ/neuron figure includes both optical and electrical energy: even though only a fraction of the energy is optical, the optical signal will be large compared to both shot noise (Eq.~(\ref{eq:nnlayer2})) and amplifier Johnson noise $\langle \Delta n_e\rangle_{\rm rms} \sim 10^{3}$ \cite{Notomi2014}, so noise will not significantly degrade the network's performance.}

\rmh{A much more aggressive goal is 1 fJ/neuron (dashed green curve).  This figure is out of reach with current technology, but research into fJ/bit on-chip interconnects may enable it in the future \cite{Notomi2014, Miller2017}.    A range of modulator designs support few-fJ/bit operation \cite{Timurdogan2014, Koos2016, Haffner2018, Srinivasan2016}.  On-chip interconnects also require photodetectors with ultra-low (fF) capacitance, so that a femtojoule of light produces a detectable signal without amplification \cite{Notomi2014, Miller2017}; such detectors have been realized with photonic crystals \cite{Nozaki2016}, plasmon antennas \cite{Ishi2005, Tang2008}, and nanowires \cite{Cao2010}.  By eliminating the amplifier, ultrasmall ``receiverless'' detectors avoid its $\sim 100$ fJ energy cost as well as the Johnson noise associated with the amplifier.  (Johnson noise still leads to fluctuations in the capacitor charge (kTC noise) that go as $\langle \Delta n_e\rangle_{\rm rms} = \sqrt{kTC}/e \approx 12 \sqrt{C/{\rm fF}}$ \cite{Pierce1956}, but for small detectors shot noise will dominate, see Supp.~Sec.~\ref{sec:s4}).  Since 1 fJ/neuron is below the energy figures for ADCs, it would require well-designed analog electronics (for the nonlinear activation function) and very tight integration between detector, logic, and modulator \cite{Miller2017}.  At these energies, shot noise is also non-negligible and} the SQL becomes relevant, but as mentioned above, due to optical inefficiencies the SQL will likely be relevant at higher energies as well.

For context, the $\sim$1 pJ/MAC figure \cite{Keckler2011, Chen2014, Jouppi2017} for state-of-the-art ASICs is shown in Fig.~\ref{fig:f3}(c).  Energy consumption in non-reversible logic gates is bounded 
by the Landauer (thermodynamic) limit $E_{\rm op} = kT\log(2) \approx 3\,{\rm zJ}$ \cite{Landauer1961}.  While multiply-and-accumulate is technically a reversible operation, all realistic computers implement it using non-reversible binary gates, so Landauer's principle applies.  A 32-bit multiplication \cite{Nagamatsu1989, Yao1993} requires approximately $10^3$ binary gates (see Supp.~Sec.~\ref{sec:s4}) and each bit operation consumes at least $kT\log(2)$, giving a limit $E_{\rm mac} \geq 3\,{\rm aJ}$ (dotted line in Fig.~\ref{fig:f3}(c)).  This is already higher than the SQL for the larger networks  with $N \geq 100$.  The optical neural network can achieve sub-Landauer performance because \rmh{(1) it operates in analog, avoiding the overhead of many bit operations per multiplication, and (2)} the matrix product is performed through optical interference, which is reversible and not subject to the bound.  \rmh{To understand the second point, recall that homodyne detection computes the dot product via the polarization identity: $\vec{u}\cdot \vec{v} = \tfrac{1}{4}(\lVert \vec{u}+\vec{v}\rVert^2 - \lVert \vec{u}-\vec{v}\rVert^2)$.  Optical interference, the reversible element that breaks Landauer's assumption, is needed to convert the signals representing $\vec{u}$ and $\vec{v}$ to $\vec{u} \pm \vec{v}$ before squaring on the detectors and subtracting.}

A final consideration is the \rmh{electrical} energy required to generate the weights.  
There is one weight pulse per MAC, so at the minimum this will be 1 fJ/MAC for the modulator, and may rise above 1 pJ/MAC once the driver electronics and memory access are included.  However, once the optical signal is generated, it can be fanned out to many neural networks in parallel, reducing this cost by a factor of $B$, the batch size.  Large batch sizes should enable this contribution to $E_{\rm mac}$ to reach the few-femtojoule regime, and potentially much lower.  

\subsection*{Training and Convolutions with Optical GEMM}

\begin{figure}[tb]
\begin{center}
\includegraphics[width=1.00\textwidth]{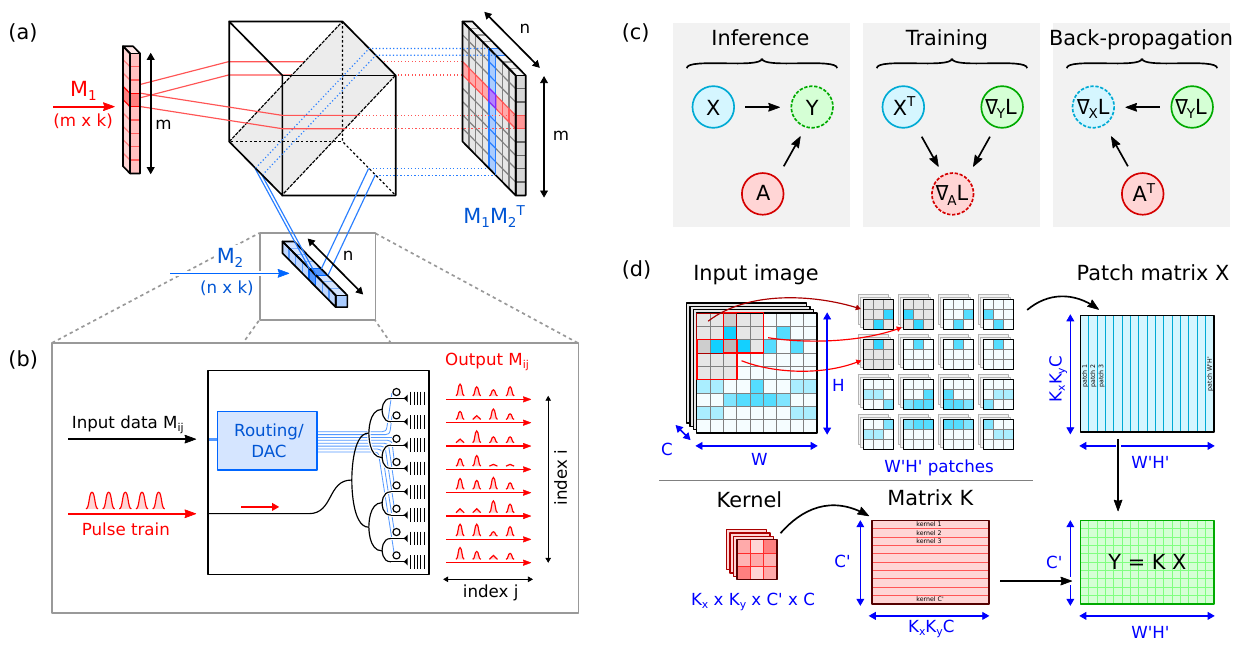}
\caption{(a) Matrix multiplication with a 2D detector array, two 1D transmitter arrays, and optical fan-out.  Imaging lenses (including cylindrical lenses for row- and column-fanout) not shown.  (b) Schematic diagram of transmitter array.  (c) Required matrix operations for inference, training, and back-propagation in a deep neural network.  (d) Patching technique to recast a convolution ($K_x = K_y = 3, s_x = s_y = 2$ shown) as a matrix-matrix multiplication.}
\label{fig:f4}
\end{center}
\end{figure}

As discussed previously, the optical unit in Fig.~\ref{fig:f1}(b) performs a matrix-vector product, and running multiple units in parallel with the same set of weights performs a general matrix-matrix product (GEMM), a key function in the Basic Linear Algebra Subprograms (BLAS) \cite{Lawson1979}.  Fig.~\ref{fig:f4}(a-b) shows a schematic for an optical GEMM unit based on homodyne detection inspired by the neural-network concept.  The inputs are two matrices $(M_1)_{m\times k}$ and $(M_2)_{n\times k}$, encoded into optical signals on the 1D red (blue) \rmh{integrated photonic} transmitter arrays.  Cylindrical lenses map these inputs to rows (columns) of the 2D detector array.  From the accumulated charge at each pixel, one can extract the matrix elements of the product $(M_1 M_2^T)_{m\times n}$.  This operation requires $m\cdot n\cdot k$ MACs, and the total energy consumption (and energy per MAC) are:
\bea
	E_{\rm tot} & = & (m k + n k) E_{\rm in} + (m n) E_{\rm out} \nonumber \\
	E_{\rm mac} & = & \left(\frac{1}{n} + \frac{1}{m}\right) E_{\rm in} + \frac{1}{k} E_{\rm out} \label{eq:gemmenergy}
\eea
where $E_{\rm in}$, $E_{\rm out}$ are the transmitter and receiver energy requirements, per symbol, which include all optical energy plus electronic driving, serialization, DAC/ADC, etc.  If all matrix dimensions $(m, n, k)$ are large, significant energy savings per MAC are possible if $E_{\rm in}, E_{\rm out}$ can be kept reasonably small.

We saw above that the optical system could be used for neural-network inference.  When running a batch of $B$ instances $X = [x_1 \ldots x_B]$, the output $Y = [y_1 \ldots y_B]$ can be computed through the matrix-matrix product $Y = A X$.  In fully-connected layers, training and back-propagation also rely heavily on GEMM.  The goal of training is to find the set of weights $A^{(k)}$ that minimize the {\it loss function} $L$, which characterizes the inaccuracy of the model.  Training typically proceeds by gradient-based methods.  Since the loss depends on the network output, we start at the final layer and work backward, a process called back-propagation \cite{Werbos1974, LeCun1998}.  At each layer, we compute the gradient $(\nabla_A L)_{ij} = \partial L/\partial A_{ij}$ from the quantity $(\nabla_Y L)_{ij} = \partial L/\partial Y_{ij}$, and propagate the derivative back to the input $(\nabla_X L)_{ij} = \partial L/\partial X_{ij}$ (Fig.~\ref{fig:f4}(c)).  These derivatives are computed from the chain rule and can be written as matrix-matrix multiplications:
\beq
	\nabla_A L = (\nabla_Y L) X^T,\ \ \ 
	\nabla_X L = A^T (\nabla_Y L)
\eeq
Once the derivative has been propagated to $\nabla_{X^{(k)}} L$ (for layer $k$) we use the chain rule to compute $\nabla_{Y^{(k-1)}} L = f'(\nabla_{X^{(k)}} L)$ and proceed to the previous layer.  In this way, we sequentially compute the derivatives $\nabla_{A^{(k)}} L$ at each layer in the neural network.

In addition to fully-connected layers, it is also possible to run convolutional layers on the optical GEMM unit by employing a ``patching'' technique \cite{Chetlur2014}.  In a convolutional layer, the input $x_{ij;k}$ is a $W\times H$ image with $C$ channels.  This is convolved to produce an output $y_{ij;k}$ of dimension $W'\times H'$ with $C'$ channels \cite{Sze2017}:
\beq
	y_{ij;k} = \sum_{i'j',l} K_{i'j',kl} x_{(s_xi+i')(s_yj+j'); l} \label{eq:conv}
\eeq
Here $K_{i'j',kl}$ is the convolution kernel, a 4-dimensional tensor of size $K_x\times K_y\times C'\times C$, and $(s_x, s_y)$ are the strides of the convolution.  Na\"{i}vely vectorizing Eq.~(\ref{eq:conv}) and running it as a fully-connected matrix-vector multiply is very inefficient because the resulting matrix is sparse and contains many redundant entries.  Patching expresses the image as a matrix $X$ of size $K_xK_y C \times W'H'$, where each column corresponds to a vectorized $K_x\times K_y$ patch of the image (Fig.~\ref{fig:f4}(d)).  The elements of the kernel are rearranged to form a (dense) matrix $K$ of size $C'\times K_xK_y C$.  Eq.~(\ref{eq:conv}) can then be computed by taking the matrix-matrix product $Y = K X$, which has size $C' \times W'H'$.  On virtually any microprocessor, GEMM is a highly optimized function with very regular patterns of memory access; the benefits of rewriting the convolution as a GEMM greatly outweigh the redundancy of data storage arising from overlapping patches \cite{Chetlur2014}.  The time required to rearrange the image as a patch matrix is typically very small compared to the time to compute the GEMM \rmh{\cite{Li2016} (and can be further reduced if necessary with network-on-chip architectures \cite{Chen2017} or optical buffering \cite{Bagherian2018})}; therefore, by accelerating the GEMM, the optical matrix multiplier will significantly improve the speed and energy efficiency of convolutional layers.  Note also that, since we are performing the convolution as a matrix-matrix (rather than matrix-vector) operation, it is possible to obtain energy savings even without running the neural network on large batches of data.  Computing the convolution requires $W'H' K_x K_y C'C$ MACs.  Following Eq.~(\ref{eq:gemmenergy}), the energy per MAC (not including memory rearrangement for patching) is:
\beq
	E_{\rm mac} = \underbrace{\left(\frac{1}{C'} + \frac{1}{W'H'}\right)}_{1/c_{\rm in}} E_{\rm in} + \underbrace{\frac{1}{K_xK_y C}}_{1/c_{\rm out}} E_{\rm out} \label{eq:convsavings}
\eeq
The coefficients $c_{\rm in} = (1/C + 1/W'H')^{-1}$ and $c_{\rm out} = K_xK_y C$ govern the energy efficiency when we are limited by input / output energies (transmitter / receiver and associated electronics).  Since reading a 32-bit register takes $\sim$pJ of energy \cite{Sze2017}, a reasonable lower bound for near-term systems is $E_{\rm in}, E_{\rm out} \gtrsim {\rm pJ}$.  Thus it is essential that $c_{\rm in}, c_{\rm out} \gg 1$ for the energy performance of the optical system to beat an ASIC ($\sim$pJ/MAC).

\begin{table}[b!]
\begin{center}
\begin{tabular}{c|cccc|c|cc}
\hline\hline
Layer & Input & Output & Kernel & Stride & MACs & $c_{\rm in}$ & $c_{\rm out}$ \\ \hline
CONV1 & $227\times 227\times 3$ &  $55\times 55\times 96$ & $11\times 11\times 96\times 3$ & 4 & 105M & 93 & 363 \\
(pool) & $55\times 55\times 96$ & $27\times 27\times 96$ & -- & 2 & -- & -- & -- \\ 
CONV2 & $27\times 27\times 96$ & $27\times 27\times 256$  & $5\times 5\times 256\times 96$ & 1 & 448M & 189 & 2400 \\
(pool) & $27\times 27\times 256$ & $13\times 13\times 256$ & -- & 2 & -- & -- & -- \\ 
CONV3 & $13\times 13\times 256$ & $13\times 13\times 384$ & $3\times 3\times 384\times 256$ & 1 & 150M & 117 & 2304 \\
CONV4 & $13\times 13\times 384$ & $13\times 13\times 384$ & $3\times 3\times 384\times 384$ & 1 & 224M & 117 & 3456 \\
CONV5 & $13\times 13\times 384$ & $13\times 13\times 256$ & $3\times 3\times 256\times 384$ & 1 & 150M & 102 & 3456 \\ (pool) & $13\times 13\times 256$ & $6\times 6\times 256$ & -- & 2 & -- & -- & -- \\ \hline
FC1 & $6\times 6\times 256$ & $4096$ & -- & -- & 38M & -- & -- \\
FC2 & $4096$ & $4096$ & -- & -- & 17M & -- & -- \\
FC2 & $4096$ & $1000$ & -- & -- & 4M & -- & -- \\ \hline
\multicolumn{5}{r|}{Total CONV layers} & 1.08G & 132 & 1656 \\
\multicolumn{5}{r|}{Total FC layers} & 59M & -- & -- \\
\hline\hline
\end{tabular}
\caption{Layers in AlexNet \cite{Krizhevsky2012}.  Values of $c_{\rm in}, c_{\rm out}$ are calculated from Eq.~(\ref{eq:convsavings}).  Max-pooling layers after CONV1, CONV2, and CONV5 are used to reduce the image size, but the relative computational cost for these layers is negligible.  }
\label{tab:t1}
\end{center}
\end{table}

As a benchmark problem, we consider AlexNet \cite{Krizhevsky2012}, the first convolutional neural network to perform competitively at the ImageNet Large-Scale Visual Recognition Challenge \cite{ILSVRC15}.  AlexNet consists of 5 convolutional (CONV) layers and 3 fully-connected (FC) layers, and consistent with deep neural networks generally, the majority of the energy consumption comes from the CONV layers \cite{Sze2017}.  Table~\ref{tab:t1} gives the layer dimensions and the values of $c_{\rm in}, c_{\rm out}$ for the CONV layers in AlexNet \cite{Krizhevsky2012}.  The MAC-weighted averages for all layers are $\langle c_{\rm in} \rangle > 100$ and $\langle c_{\rm out} \rangle > 1000$.  Thus, even under extremely conservative assumptions of $E_{\rm in}, E_{\rm out} \gtrsim 100\,{\rm pJ}$ (comparable to DRAM read energies \cite{Sze2017, Horowitz2014}), it is still possible to achieve sub-pJ/MAC performance.

\begin{figure}[tbp]
\begin{center}
\includegraphics[width=1.00\textwidth]{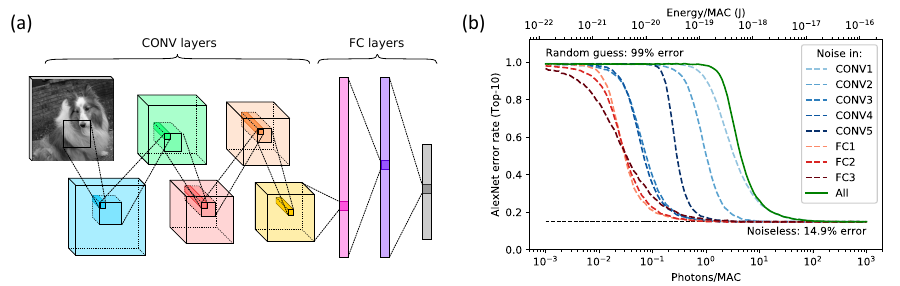}
\caption{(a) Schematic drawing of AlexNet, which consists of 5 convolutional layers and 3 fully-connected layers.  Pooling and normalization steps not shown.  (b) Error rate for pre-trained AlexNet as a function of $n_{\rm mac}$.  Dashed lines show the effect of noise in a single layer, while solid green line shows the performance of the actual machine where all layers have noise.}
\label{fig:f5}
\end{center}
\end{figure}

More advanced technology, such as few-fJ optical interconnects \cite{Miller2017}, may significantly reduce $E_{\rm in}$ and $E_{\rm out}$, and therefore the energy per MAC.  However, the performance is still fundamentally limited by detector shot noise (e.g. Eq.~(\ref{eq:nnlayer2}) for FC layers).  Supp.~Sec.~\ref{sec:s3} extends the shot-noise analysis to the case of matrix-matrix products needed for the convolutional case.  Using a pre-trained AlexNet model (see Methods for details), Fig.~\ref{fig:f5}(b) shows the top-10 accuracy on the ImageNet validation set as a function of the number of photons per MAC $n_{\rm mac}$.  Consistent with Fig.~\ref{fig:f2}(c), there are two limits: $n_{\rm mac} \ll 1$ corresponds to the random guess regime with 99\% error rate (for top-10 accuracy with 1,000 classes), while $n_{\rm mac} \gg 1$ recovers the accuracy of the noiseless model.

The dashed lines in Fig.~\ref{fig:f5}(b) show the fictitious case where noise is present in only a single layer, while the solid green line corresponds to the case where all layers have noise and $n_{\rm mac}$ is the same for each layer.  Not all layers contribute equally to the noise: CONV1 is the most sensitive, requiring $n_{\rm mac} \gtrsim 20$, while the deeper layers (particularly the fully-connected layers) can tolerate much lower energies $n_{\rm mac} \gtrsim 1$.  Since the SNR is related to the total power received, which scales as $c_{\rm out} n_{\rm mac}$ for the convolutional layers ($c_{\rm out}$ pulses per detector), it is not surprising that the deeper layers, which have a larger $c_{\rm out}$, are less sensitive to quantum noise.  The SQL obtained for AlexNet ($n_{\rm mac} \gtrsim 20$ or $E_{\rm mac} \gtrsim 3\,$aJ) is slightly larger than that from the MNIST networks in Fig.~\ref{fig:f2}(c), but of the same order of magnitude, suggesting that the SQL is somewhat problem-dependent.

\rmh{It is worth contrasting the optical GEMM to more familiar optical convolvers.  It has long been known that 2D convolutions can be performed with optical Fourier transforms \cite{Lugt1964, Paek1987, New2017}.  However, this technique suffers from two significant drawbacks.  First, it employs spatial light modulators, which limits the speed at which the kernel can be reprogrammed.  In addition, optics performs a single-channel ($C = C' = 1$) convolution, and while extending to multiple output channels is possible by tiling kernels \cite{Chang2018}, multiple input {\it and} output channels may be difficult.}

\rmh{In contrast to free-space and fully-integrated approaches, the optical GEMM leverages the complementary strengths of both free-space and integrated photonics.  Integrated photonics is an ideal platform for realizing the transmitters, as these employ a large number of fast (GHz) modulators on chip.  On-chip integration allows scaling to large arrays with control over the relative phase of each output beam (a capability exploited in recent chip-based phased arrays for beam steering \cite{Sun2013, Chung2018, Phare2018}).  Free-space propagation provides an essential third dimension, which enables high bandwidths at moderate clock frequencies \cite{Miller2017} and data fan-out patterns that are difficult to implement on a 2D photonic chip.  However, having a free-space element leads to issues with phase stability and aberrations.  Since the transmitters are integrated, it is the relative phase between the beam paths that drifts (on timescales long compared to a computation), and this can be stabilized with a single feedback loop to the overall transmitter phase, a small constant overhead that does not scale with matrix size.  To correct for geometric aberrations and minimize crosstalk between detectors, multi-lens arrangements can be used, a standard practice in high-resolution imaging systems \cite{Smith1966}.  Supp.~Sec.~\ref{sec:s2} presents an optical design and analysis using Zemax\textregistered{} simulation software supporting the hypothesis that a $10^3 \times 10^3$ optical GEMM is achievable.}

\subsection*{Discussion}

This paper has presented a new architecture for optically accelerated deep learning that is scalable to large problems and can operate at high speeds with low energy consumption.  Our approach takes advantage of the photoelectric effect, via the relation $I \propto |E|^2$, to compute the required matrix products opto-electronically \rmh{without need for an all-optical nonlinearity, a key difficulty that has hobbled conventional approaches to optical computing \cite{Miller2010b}}.  Since the device can be constructed with free-space optical components, it can scale to much larger sizes than \rmh{purely} nanophotonic implementations \cite{Shen2017}, being ultimately limited by the size of the detector array ($N \gtrsim 10^6$).

A key advantage to this scheme is that the multiplication itself is performed passively by optical interference, so the main speed and energy costs are associated with routing data into and out of the device.  For a matrix multiplication $C_{m\times n} = A_{m\times k} B_{k\times n}$, the input/output (I/O) energy scales as $O(mk) + O(nk) + O(mn)$, while the number of MACs scales as $O(mnk)$.  For moderately large problems found in convolutional neural-network layers ($m, n, k \geq 100$) with moderate I/O energies ($\sim$pJ), performance in the $\sim$10 fJ/MAC range should be feasible, which is 2--3 orders of magnitude smaller than for state-of-the-art CMOS circuits \cite{Keckler2011, Chen2014, Jouppi2017}.  Advances in optical interconnects \cite{Timurdogan2014, Sun2015, Atabaki2018} may reduce the I/O energies by large factors \cite{Miller2017}, translating to further improvements in energy per MAC.

The fundamental limits to a technology are important to its long-term scaling.  For the optical neural network presented here, detector shot noise presents a standard quantum limit to neural network energy efficiency \cite{Caves1981}.  Because this limit is physics-based, it cannot be engineered away unless non-classical states of light are employed \cite{Jaekel1990, Grote2013}.  To study the SQL in neural networks, we performed Monte Carlo simulations on pre-trained models for MNIST digit recognition (fully-connected) and ImageNet image classification (convolutional).  In both cases, network performance is a function of the number of photons used, which sets a lower bound on the energy per MAC.  This bound is problem- and network-dependent, and for the problems tested in this paper, lies in the range $50\,{\rm zJ}$--$5\,{\rm aJ}$/MAC.  By contrast, the Landauer (thermodynamic) limit \rmh{for a digital processor} is 3 aJ/MAC (assuming 1,000 bit operations per MAC \cite{Nagamatsu1989, Yao1993}); sub-Laudauer performance is possible because the multiplication is performed through optical interference, which is reversible and not bounded by Landauer's principle.

Historically, the exponential growth in computing power has driven advances in machine learning by enabling the development of larger, deeper, and more complex models \cite{Steinkraus2005, Sze2017, Chen2014, Jouppi2017}.  As Moore's Law runs out of steam, photonics may become necessary for continued growth in processing power---not only for interconnects \cite{Miller2017}, but also for logic.  The architecture sketched in this paper promises significant short-term performance gains over state-of-the-art electronics, with a long-term potential, bounded by the standard quantum limit, of many orders of magnitude of improvement.

\subsection*{Methods}

Neural-network performance was computed using Monte Carlo simulations.  For fully-connected layers, Eq.~(\ref{eq:nnlayer2}) was used, while for convolutional layers, the convolution was performed by first forming the patch matrix (Fig.~\ref{fig:f4}(d)) and performing the matrix-matrix multiplication (noise model discussed in Supp.~Sec.~\ref{sec:s3}).  The weights for the fully-connected MNIST neural networks were trained on a GPU using TensorFlow.  A pretrained TensorFlow version of AlexNet (available online at Ref.~\cite{Peng2018}) was modified to implement the quantum noise model and used for ImageNet classification.  Simulations were performed on an NVIDIA Tesla K40 GPU.

\subsection*{Acknowledgements}

R.H.\ is supported by an IC Postdoctoral Research Fellowship at MIT, administered by ORISE through U.S.\ DOE and ODNI.  L.B.\ is supported by a Doctoral Postgraduate Scholarship from the Natural Sciences and Engineering Research Council of Canada (NSERC).  D.E.\ and M.S.\ acknowledge support from the U.S.~ARO through the ISN at MIT (no.~W911NF-18-2-0048).  The authors acknowledge John Peurifoy (MIT) for training a number of the MNIST neural-network models, and NVIDIA Corporation for the donation of the Tesla K40 GPU used in this research.  We are grateful to Joel Emer (NVIDIA / MIT) and Vivienne Sze (MIT) for helpful discussions.

Correspondence and requests for materials should be addressed to R.H.~(email: rhamerly@mit.edu).


\newpage

\section*{[Supplementary] Large-Scale Optical Neural Networks based on Photoelectric Multiplication}
\renewcommand{\thetable}{S\arabic{table}}
\renewcommand{\thefigure}{S\arabic{figure}}
\renewcommand{\thesection}{S\arabic{section}}
\renewcommand{\thesubsection}{S\arabic{section}.\arabic{subsection}}
\renewcommand{\theequation}{S\arabic{equation}}
\setcounter{section}{0}
\setcounter{figure}{0}
\setcounter{table}{0}
\setcounter{equation}{0}

\section{Homodyne Product Implementation Details}
\label{sec:s1}

\begin{figure}[b!]
\begin{center}
\includegraphics[width=0.95\textwidth]{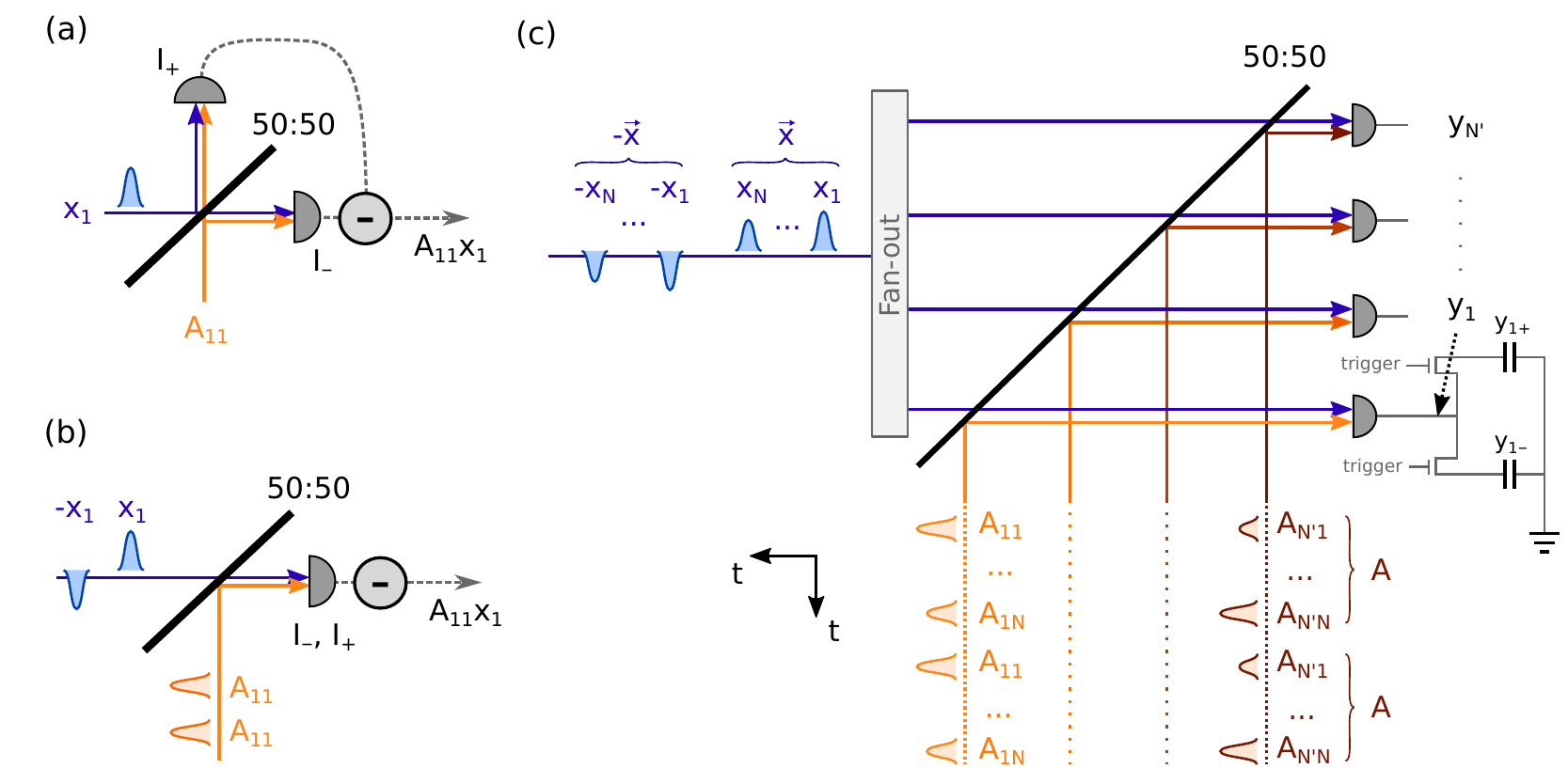}
\caption{Schematic diagram of homodyne detection for one layer of the optical neural network. (a) Standard homodyne detection. (b) Homodyne detection with a single detector, components encoded sequentially in time. (c) Full schematic of homodyne detection for a single layer of the neural network.}
\label{fig:fs1}
\end{center}
\end{figure}

Using optical homodyne detection (Fig.~\ref{fig:fs1}(a)), it is possible to obtain a signal proportional to the product of electric field amplitudes originating from two coherent, in-phase optical sources at different spatial locations. The input beams of electric field amplitudes $x_{1}$ and $A_{11}$ travel through a 50:50 beamsplitter, interfere, and outputs $I_{+}$ and $I_{-}$ are detected, where $I_{+}=\tfrac{1}{2}|x_1 + A_{11}|^2$ and $I_{-}=\tfrac{1}{2}|x_1 - A_{11}|^2$. The difference of the photocurrents $I_{+}-I_{-}$ is proportional to the real part of the product of the incident electric field amplitudes $2\mbox{Re}[A_{11}^* x_{1}]$.  If all field amplitudes are real, this returns the product $2A_{11} x_1$.

Alternatively, to reduce system complexity and minimize data transfer requirements, we can instead opt to use a single detector to perform homodyne detection (Fig.~\ref{fig:fs1}(b)). In this case, two copies of the signal $x_1$ and weight $A_{11}$ are sent into the detector, and a phase modulator applies a $\pi$-phase shift to the second copy of $x_1$, flipping its amplitude.  The photocurrents $I_+$ and $I_-$ now appear separated in time rather than space, and are read out separately and may be subsequently subtracted.  This technique can be applied over many channels and time steps, such that the full matrix-vector product $\vec{y} = A\vec{x}$ is computed (Fig.~\ref{fig:fs1}(c)).  In this technique, half of the light is discarded and twice as many pulses are required, doubling the energy consumption and latency.  However, since the same detector and beamsplitter are used for $I_+$ and $I_-$, one avoids the technical issues associated with  beamsplitter imbalance and detector inhomogeneities.

\begin{figure}[tb]
\begin{center}
\includegraphics[width=0.65\textwidth]{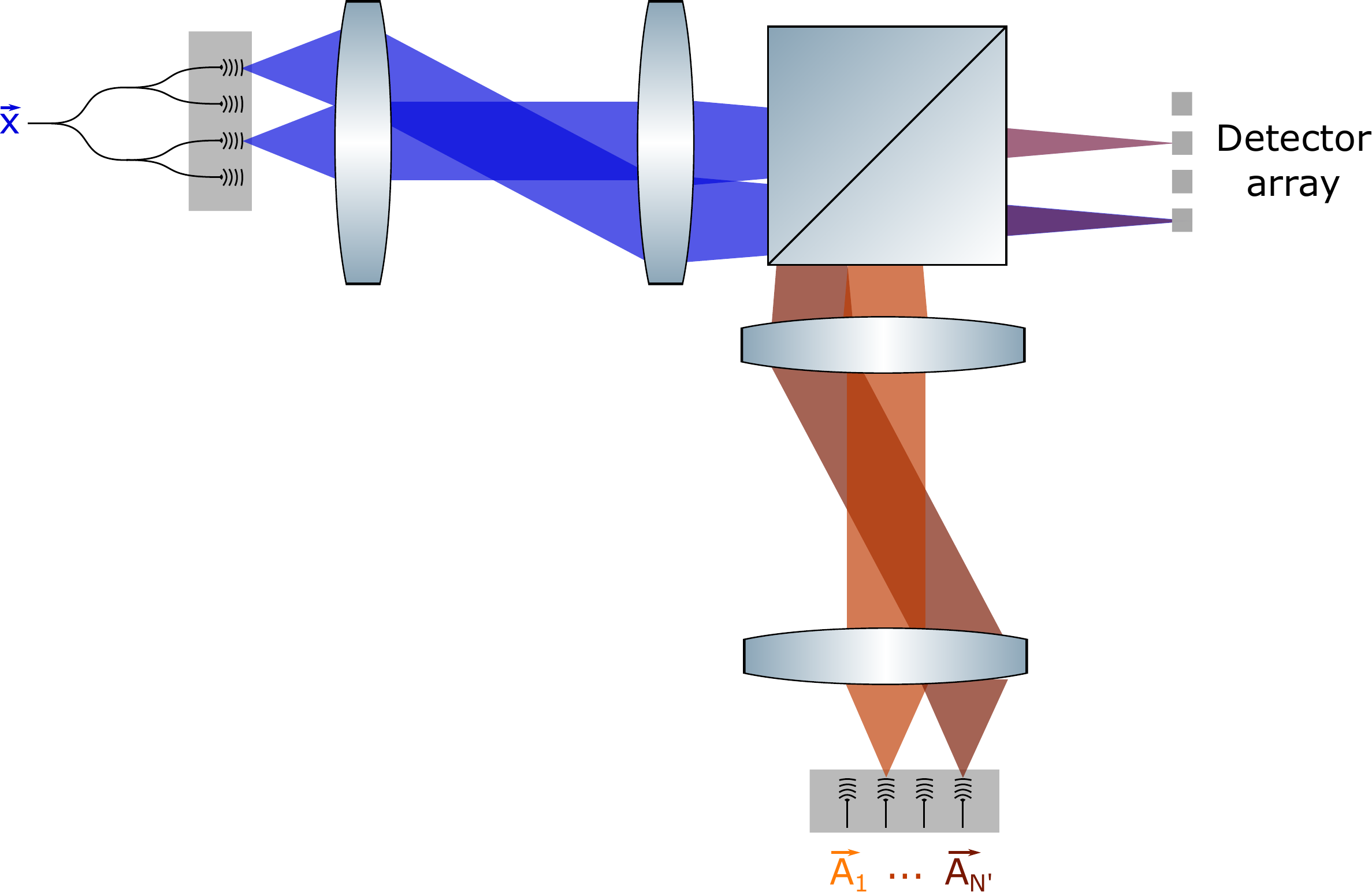}
\caption{Experimental implementation of homodyne detection for optical neural network. $\vec{x}$ is fanned out and the weight-matrix $A$ is generated on-chip. Outcoupling is achieved with grating antennas or nanoantennas, beams are focused onto separate detector pixels. Two example beams are drawn.}
\label{fig:fs2}
\end{center}
\end{figure}

In practice, optical fan-out of $\vec{x}$ and encoding of the weight-matrix $A$ by intensity and phase modulation can be performed on-chip. Optical phased arrays containing as many as 1024 grating antennas for outcoupling have been shown in a sub-cm$^2$ CMOS chip, including 1192 phase and 168 amplitude modulators \cite{Chung2018}. The signals can be interfered and imaged on the detector array using bulk optics. An example experimental setup is illustrated in Fig.~\ref{fig:fs2}.

\rmh{The pulse trains for both $x_i$ and $A_{ij}$ must come from the same master laser, which can be either a pulsed laser or a continuous-wave laser that passes through an intensity modulator.  For high-performance systems, moderate powers are required (e.g.\ assuming 0.1 pJ optical energy per detector, a 1-GHz $1000\times 1000$ optical GEMM will require $100\,{\rm mW}$ of laser power), so an optical amplifier may be needed.  The path length difference must be less than both (1) the coherence length of the laser and (2) $c\tau_{\rm pulse}$.  Although these conditions are not very stringent (assuming $\tau_{\rm pulse} \gtrsim 100\,{\rm ps}$), using a design with equal path lengths for data and weights is straightforward in practice.}

\section{Aberration Management}
\label{sec:s2}

\begin{figure}[b!]
\begin{center}
\includegraphics[width=0.9\textwidth]{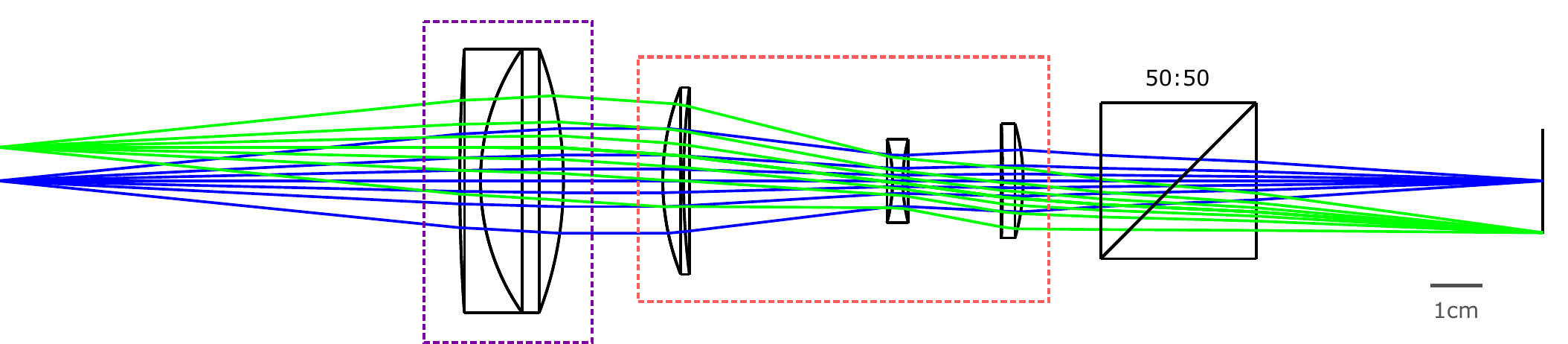}
\caption{\rmh{Example design of free-space portion of optical neural network optimized with Zemax\textregistered{}.  Actual apparatus will have two arms, as shown in Fig.~\ref{fig:fs2}.  On-axis propagation shown in blue, off-axis beam furthest from center shown in green.  Source array of height $-6.6\,{\rm mm}$ to $+6.6\,{\rm mm}$ sends light through Thorlabs achromat (AC508-100-A, ${\rm f}=100\,{\rm mm}$, purple box), then through a custom optimized lens triplet (${\rm f}=150\,{\rm mm}$, red box) and finally through a 50:50 beamsplitter before reaching detector.  Lens triplet can be spherical (for matrix-vector multiplication) or cylindrical (matrix-matrix multiplication).}}
\label{fig:fs2_1}
\end{center}
\end{figure}

\lb{Optical aberrations are a possible limit to scalability of our free-space homodyne optical neural network.  In order to assess their impact, we used Zemax\textregistered{} ray-tracing software to model and optimize the lower branch of Fig.~\ref{fig:fs2}, which provides the weights $A_{ij}$.  Specifically, on- and off-axis sources were propagated through an off-the-shelf achromatic lens (Thorlabs AC508-100-A) for collimation, a custom Cooke triplet of spherical lenses for focusing, and a cube beamsplitter.  To achieve this design, illustrated in Fig.~\ref{fig:fs2_1}, we varied lens spacing and radii of curvature to minimize wavefront distortion due to Seidel aberrations across a $20\,{\rm mm}\times 20\,{\rm mm}$ field of view.  To focus the beam in one dimension and collimate it in the other (required for the optical GEMM unit proposed in~Fig.~6), the three circular lenses used for focusing can be replaced with cylindrical lenses with the same thicknesses and radii of curvature.}

\begin{figure}[t!]
\begin{center}
\includegraphics[width=0.9\textwidth]{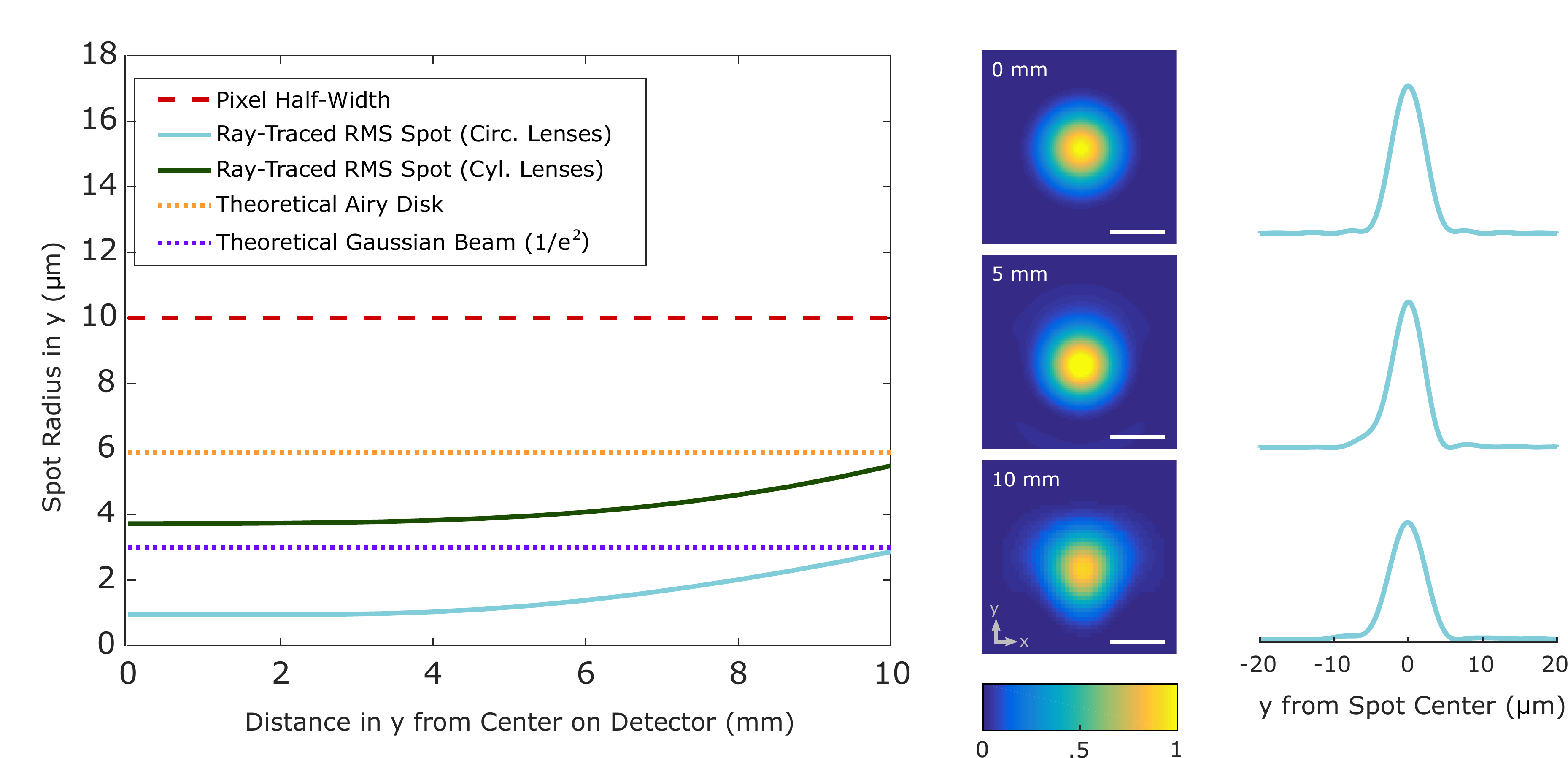}
\caption{\rmh{Left: simulated spot radius as a function of distance from center of detector (light emitted from different sources reaches different detector pixels).  RMS spot radius of traced rays and Airy disk radius calculated in Zemax\textregistered{}.  Gaussian beam radius calculated with Gaussian beam theory.  Right: beam shape at $y = 0$, $5$, and $10\,{\rm mm}$ from center of detector simulated using Physical Optics module in Zemax\textregistered{} shown in 2D colormap and 1D (at $x = 0$).  Scale bars: $5\,{\rm \mu m}$.  Colorbar: irradiance, linear arb. units.}} 
\label{fig:fs2_2}
\end{center}
\end{figure}

\lb{In the system described and modeled here, 1000 sources at $\lambda =635\,{\rm nm}$, each with a mode field diameter of $4\,{\rm \mu m}$ and separated by $13\,{\rm \mu m}$, are imaged onto a detector with one million $20\,{\rm \mu m}\times 20\,{\rm \mu m}$ pixels. The active area of each pixel is confined to the center $5\,{\rm \mu m}\times 5\,{\rm \mu m}$ to minimize crosstalk from neighboring pixels. As shown in Fig.~\ref{fig:fs2_2}, the Zemax\textregistered{} ray-traced aberrated spot sizes on the detector were found to be below the diffraction limit, and Zemax\textregistered{} ``Physical Optics" simulations (based on scalar diffraction theory) confirm that we achieve small, near-Gaussian focused spots over the entire field of view.  The substitution of circular lenses with cylindrical lenses does not lead to a significant decrease in lens performance, as the RMS spot size (in the dimension being focused) remains small.}

\lb{The ``Physical Optics" simulations were also used to estimate crosstalk, defined as undesired light intensity from adjacent pixels divided by signal intensity on the pixel's active area.  We found that \textgreater50\% of the optical energy is detected in the active area, with $\sim$1\% crosstalk from 8 nearest neighbors over the entire field of view.  Interestingly, most of the crosstalk on the detector edges is due to a single neighbor, as the spot is aberrated asymmetrically.}

\rmh{Because of the phase coherence between the neighboring spots, constructive interference between a spot and its neighbors may make the actual crosstalk larger than the intensity figure quoted above.}  \lb{If all the light on the detector is assumed to be in phase (worst case), simulations show the crosstalk increases to $\sim$2\%.  If this level of crosstalk is too high for an application, it can be precalculated and compensated (to first order) in logic in the transmitter array.}  \rmh{Based on these results, optical networks with up to $10^6$ channels should be achievable with reasonable signal fidelity.  Further engineering improvements, e.g.\ apodizing the Airy disk with a Hann window, may lead to reduced crosstalk and larger system size limits.}

\section{Shot Noise in Neural-Network Layers}
\label{sec:s3}

\subsection{Matrix-Vector Multiply}

\label{sec:supp-mv}

In the time-encoded neural network, at each layer, a stream of data $\bar{x}_i$ is broadcast to the neurons of the subsequent layer.  Each neuron is a homodyne detector that interferes this broadcast signal against the weight signal $\bar{A}_{ij}$.  Assume perfect spatial and temporal mode-matching between input and weight signals, and normalize the quantities so that $|\bar{x}_i|^2$, $|\bar{A}_{ij}|^2$ correspond to the number of photons per pulse.  If a pulse with amplitude $\bar{u}$ enters the detector, the output current is Poisson distributed:
\beq
	\frac{Q}{e} \sim {\rm Poisson}(|u|^2)
\eeq
This has a mean of $|u|^2$ and a standard deviation of $|u|$.  The homodyne signal at neuron $i$ is obtained by interfering the signals $\bar{x}_j$ and $\bar{A}_{ij}$ on a 50/50 beamsplitter and taking the difference between the photocurrents $y_i = \tfrac{1}{e}(Q_i^{(+)} - Q_i^{(-)})$.  Each photocurrent $Q_i^{(\pm)}$ is the sum of many Poisson random variables, which is itself a Poisson process.  In the useful limit of many photons per neuron (though not necessarily per MAC), this is approximately Gaussian:
\beq
\frac{Q_i^{(\pm)}}{e} = \sum_j \tfrac{1}{2}(\bar{A}_{ij} \pm \bar{x}_j)^2 + w_i^{(\pm)} \Bigl(\sum_j \tfrac{1}{2}(\bar{A}_{ij} \pm \bar{x}_j)^2 \Bigr)^{1/2}
\eeq
where $w_i^{(\pm)} \sim N(0, 1)$ is a (normal) random variable.  This term is responsible for the detector shot noise.  The homodyne detector measures the difference between the photocurrents.  The means subtract while the noises add in quadrature:
\bea
	\bar{y}_i & = & \biggl[ \sum_j \tfrac{1}{2}(\bar{A}_{ij} + \bar{x}_j)^2 + w_i^{(+)} \Bigl(\sum_j \tfrac{1}{2}(\bar{A}_{ij} + \bar{x}_j)^2 \Bigr)^{1/2}\biggr]
	- \biggl[ \sum_j \tfrac{1}{2}(\bar{A}_{ij} - \bar{x}_j)^2 + w_i^{(-)} \Bigl(\sum_j \tfrac{1}{2}(\bar{A}_{ij} - \bar{x}_j)^2 \Bigr)^{1/2}\biggr] \nonumber \\
	& = & 2\sum_j \bar{A}_{ij} \bar{x}_j + w_i \Bigl(\sum_j {(\bar{A}_{ij}^2 + \bar{x}_j^2)}\Bigr)^{1/2} \nonumber \\
	& = & 2\bar{A}_{ij} \bar{x}_j + w_i \sqrt{\norm{\bar{A}_i}^2 + \norm{\bar{x}}^2} \label{eq:yi}
\eea
In the final line of (\ref{eq:yi}), the sum over $j$ is implicit, and we have used the $L_2$ vector norms for $\bar{x}$ and the row-vectors of $\bar{A}_i$ to simplify the notation as follows: $\norm{u}^2 = \sum_j |u_j|^2$.  As before, $w_i^{(\pm)}$ and $w_i$ are normally distributed as $N(0, 1)$.

This output is sent through a nonlinear function (here a ReLU) to obtain the inputs to the next layer:
\beq
	\bar{x}'_i = \alpha\, \mbox{ReLU}(\bar{y}_i^{(m)}) \label{eq:supp-sub1}
\eeq
Here $\bar{x}$, $\bar{x}'$ and $\bar{A}$ are the {\it physical} inputs/outputs and weights, normalized to the single-photon level (so that $|\bar{x}_i|^2$ is the photon number in pulse $x_i$, etc.).  These are related to the logical variables by scaling constants $\xi_x$, $\xi_A$:
\beq
	\bar{x} = \xi_x x,\ \ \ \bar{x}' = \xi_x x',\ \ \ \bar{A} = \xi_A A \label{eq:supp-sub2}
\eeq
Because the ReLU is scale-free ($\mbox{ReLU}(c x) = c\,\mbox{ReLU}(x)$ for $c > 0$), the relation between feature vectors at subsequent layers is:
\beq
	x'_i = 2\alpha\xi_A\, \mbox{ReLU}\biggl[ A_{ij} x_j + \frac{w_i}{2} \sqrt{\frac{\norm{A_{i}}^2}{\xi_x^2} + \frac{\norm{x}^2}{\xi_A^2}} \biggr] \label{eq:xinc}
\eeq
Set $\alpha = 1/(2\xi_A)$ to make the classical term match the desired relation $x'_i = \mbox{ReLU}(A_{ij} x_j)$.  The quantum noise enters inversely in $\xi_x, \xi_A$, which is expected because these quantities are related to the photon number, through the scaling of $(\bar{x}, \bar{x}', \bar{A})$.

The total number of photons involved in this layer is sum of the input photons and weight photons $n_{\rm tot} = n_{\rm tot}^{(x)} + n_{\rm tot}^{(A)}$, each given by:
\beq
	n_{\rm tot}^{(x)} = N' \norm{x}^2 \xi_x^2,\ \ \
	n_{\rm tot}^{(A)} = \norm{A}^2 \xi_A^2
\eeq
A more computationally relevant figure is the number of photons per MAC $n_{\rm mac} = n_{\rm tot}/NN'$, which can also be divided into input and weight photons $n_{\rm mac} = n_{\rm mac}^{(x)} + n_{\rm mac}^{(A)}$:
\beq
	n_{\rm mac}^{(x)} = \frac{\norm{x}^2}{N} \xi_x^2,\ \ \
	n_{\rm mac}^{(A)} = \frac{\norm{A}^2}{NN'} \xi_A^2 \label{eq:nmac}
\eeq
It is often appropriate to assume that the row vectors $A_i$ have approximately the same norm: $\norm{A_i} \approx \norm{A_j}$.  In this case, we can replace $\norm{A_i} \rightarrow \norm{A}/\sqrt{N'}$.  Making this approximation and substituting $\xi_x$, $\xi_A$ from Eq.~(\ref{eq:nmac}), we find:
\beq
	x'_i = \mbox{ReLU}\biggl[ A_{ij} x_j + w_i \frac{\norm{A}\norm{x}}{2\sqrt{NN'}} \sqrt{\frac{1}{n_{\rm mac}^{(x)}} + \frac{1}{n_{\rm mac}^{(A)}}} \biggr] \label{eq:xinc2}
\eeq
For a fixed energy per MAC $n_{\rm mac} = n_{\rm mac}^{(x)} + n_{\rm mac}^{(A)}$, the SNR is maximized when $n_{\rm mac}^{(x)} = n_{\rm mac}^{(A)} = \tfrac{1}{2} n_{\rm mac}$, which gives the form used in the main text:
\beq
	x'_i = \mbox{ReLU}\biggl[ A_{ij} x_j + w_i \frac{\norm{A}\norm{x}}{\sqrt{N^2N'}} \frac{\sqrt{N}}{\sqrt{n_{\rm mac}}} \biggr] \label{eq:xinc3}
\eeq

\subsection{Matrix-Matrix Multiply}

Fig.~6(a) of the main text introduced a modified scheme that multiplies two matrices rather than a matrix and a vector.  This is advantageous when running the network on batches of data.  It is also well-adapted to computing convolutions through the patching method (see main text).  To compute the vector-vector product $C_{m\times n} = A_{m\times k} B_{k\times n}$, pixel $(i, j)$ receives the homodyne product of two pulse trains, encoding $A_{i,:}$ (the $i^{\rm th}$ row of $A$) and $B_{:,j}$ (the $j^{\rm th}$ column of $B$).  The result is $C_{ij} = \sum_l A_{il} B_{lj}$, assuming all matrices are real.

As in Sec.~\ref{sec:supp-mv}, we normalize all quantities $\bar{A}_{ij}$, $\bar{B}_{ij}$, $\bar{C}_{ij}$ to photon number.  Following Eq.~(\ref{eq:yi}), the measured value of $\bar{C}_{ij}$ is photocurrent at detector $(i, j)$:
\beq
	\bar{C}_{ij} = 2 \sum_l \bar{A}_{il} \bar{B}_{lj} + w_{ij} \sqrt{\norm{\bar{A}_{i,:}}^2 + \norm{\bar{B}_{:,j}}^2}
\eeq
The physical quantities are related to their logical values by a scaling factor: $\bar{A} = \xi_A A, \bar{B} = \xi_B B, \bar{C} = \xi_C C$.  Scaling to logical quantities and setting $\xi_C = 2\xi_A \xi_B$ (to scale $C$ to satisfy $C = AB$ in the classical limit) we obtain:
\beq
	C_{ij} = \sum_l A_{il} B_{lj} + \frac{w_{ij}}{2} \sqrt{\frac{\norm{A_{i,:}}^2}{\xi_A^2} + \frac{\norm{B_{:,j}}^2}{\xi_B^2}} \label{eq:cij1}
\eeq
The optical energy per MAC is $n_{\rm mac} = n_{\rm mac}^{(A)} + n_{\rm mac}^{(B)}$, where $n_{\rm mac}^{(A)} = \norm{A}^2 \xi_A^2 / mk$, $n_{\rm mac}^{(B)} = \norm{B}^2 \xi_B^2 / nk$.  This allows us to trade $\xi_A, \xi_B$ for $n_{\rm mac}^{(A)}, n_{\rm mac}^{(B)}$ in Eq.~(\ref{eq:cij1}).  Also, assuming that the rows of $A$ (and columns of $B$) have roughly the same norm, we can replace $\norm{A_{i,:}} \rightarrow \norm{A}/\sqrt{m}$ and $\norm{B_{:,j}} \rightarrow \norm{B}/\sqrt{n}$.  Eq.~(\ref{eq:cij1}) then simplifies to:
\beq
	C_{ij} = \sum_l A_{il}B_{lj} + \frac{w_{ij}}{2} \frac{\norm{A}\norm{B}}{\sqrt{mnk}} \sqrt{\frac{1}{n_{\rm mac}^{(A)}} + \frac{1}{n_{\rm mac}^{(B)}}}
\eeq
As before, subject to a fixed energy constraint on $n_{\rm mac}$, setting $n_{\rm mac}^{(A)} = n_{\rm mac}^{(B)}$ maximizes the SNR.  Under this optimized choice of parameters, the input-output relation becomes:
\beq
	C_{ij} = \sum_l A_{il}B_{lj} + w_{ij} \frac{\norm{A}\norm{B}}{\sqrt{mnk}} \frac{1}{\sqrt{n_{\rm mac}}}
\eeq

\section{Johnson Noise}
\label{sec:s4}

\rmh{Johnson-Nyquist noise will lead to thermal fluctuations in measured photocurrent, which imposes its own limit on energy consumption.  However, unlike shot noise, this is hardware-dependent and can be mitigated by judicious detector design.  This section modifies the analysis of Sec.~\ref{sec:supp-mv} to include Johnson noise and determines the conditions under which it can be neglected.}

\rmh{Unlike shot noise, the amplitude of Johnson noise does not depend on the photocurrent.  Thus we modify Eq.~(\ref{eq:yi}) to include a constant noise term:}
\rmh{\beq
	 \bar{y}_i = 2\bar{A}_{ij} \bar{x}_j + w_i \sqrt{\norm{\bar{A}_i}^2 + \norm{\bar{x}}^2 + 2\langle \Delta n_e^2 \rangle}
\eeq}
\rmh{Here $\langle \Delta n_e^2 \rangle$ is the electron count variance due to thermal noise.  Noise from the two detectors adds in quadrature, leading to the factor of 2.  Making the substitutions (\ref{eq:supp-sub1}-\ref{eq:supp-sub2}, \ref{eq:nmac}), and assuming $\norm{A_i} \approx \norm{A}/\sqrt{N'}$ as above, we arrive at the formula:\beq
	x'_i = \mbox{ReLU}\biggl[ A_{ij} x_j + w_i \frac{\norm{A}\norm{x}}{2\sqrt{NN'}} \sqrt{\frac{1}{n_{\rm mac}^{(x)}} + \frac{1}{n_{\rm mac}^{(A)}} + \frac{2\langle \Delta n_e^2 \rangle}{N n_{\rm mac}^{(A)} n_{\rm mac}^{(x)}}} \biggr]
\eeq
As before, we maximize the SNR subject to the constraint $n_{\rm mac}^{(A)} + n_{\rm mac}^{(x)} = n_{\rm mac}$.  The optimal choice is the same found previously: $n_{\rm mac}^{(A)} = n_{\rm mac}^{(x)} = \tfrac{1}{2} n_{\rm mac}$.  This leads to\beq
	x'_i = \mbox{ReLU}\biggl[ A_{ij} x_j + w_i \frac{\norm{A}\norm{x}}{\sqrt{NN'}} \frac{1}{\sqrt{n_{\rm mac}}} \sqrt{1 + \frac{2 \langle \Delta n_e^2 \rangle}{N n_{\rm mac}}} \biggr]
\eeq
This differs from Eq.~(\ref{eq:xinc3}) by the factor $\sqrt{1 + 2 \langle \Delta n_e^2\rangle /N n_{\rm mac}}$ in the noise term.  Johnson noise is dominant when $N n_{\rm mac} \ll 2 \langle \Delta n_e^2\rangle$, and shot noise is dominant when $N n_{\rm mac} \gg 2\langle\Delta n_e^2\rangle$.  The value of $\langle \Delta n_e^2\rangle$ depends on the detector design.  For typical optical receivers with amplifiers, it can be quite large.  For femtofarad ``receiverless'' detectors being studied for on-chip interconnects, one can estimate $\langle \Delta n_e^2\rangle$ from the capacitor kTC formula $\langle \Delta n_e^2 \rangle = kTC/e^2$ \cite{Pierce1956}.  The ratio of Johnson noise to shot noise at the SQL is thus:\beq
	\frac{\langle \Delta n^2\rangle_{\rm th}}{\langle \Delta n^2\rangle_{\rm shot}} = \frac{2 \langle \Delta n_e^2 \rangle}{N n_{\rm mac}} = \frac{C}{N n_{\rm mac} e^2/2kT} \equiv \frac{C}{C_0}
\eeq
where we have defined $C_0 \equiv N n_{\rm mac} e^2/2kT$.  $C_0$ is a measure of how small a detector's capacitance must be to achieve a quantum-limited (rather than thermally limited) performance, and depends on the neural network layer.  Values for the networks studied in Sec.~II are calculated in Table \ref{tab:ts2}.  From these figures, we see that $C_0$ is in the few-femtofarad range, sub-femtofarad receiverless detectors should be able to reach the SQL, while larger detectors will be limited by Johnson noise.}

\begin{table}[b!]
\begin{center}
\begin{tabular}{l|c|cc|c|c}
\hline\hline
Model & SQL $n_{\rm mac}$ & Layer & $N$ & $N n_{\rm mac}$ & $C_0 = N n_{\rm mac} e^2/2kT$ \\ \hline
         &     & 1 & 768  & 3840--7680 & 12--24 fF \\
Small NN & 5--10   & 2 & 100  & 500--1000 & 1.6--3.2 fF \\
         &     & 3 & 100  & 500--1000  & 1.6--3.2 fF \\ \hline
         &     & 1 & 768  & 384--768  & 1.2--2.4 fF \\
Large NN & 0.5--1 & 2 & 1000 & 500--1000 & 1.6--3.2 fF \\
         &     & 3 & 1000 & 500--1000  & 1.6--3.2 fF \\
\hline\hline
\end{tabular}
\caption{\rmh{Calculation of $C_0$ for neural networks introduced in Sec.~II.}}
\label{tab:ts2}
\end{center}
\end{table}

\rmh{By going to cryogenic temperatures, it is possible to reduce the Johnson noise and thus the on-chip operating power.  However, the noise-limited power scales linearly as $P \propto T_c$ ($T_c$ is the chip temperature), so the heat dissipated goes as $dQ_c \propto T_c$ and the entropy created in the chip $dS = dQ/T_c$ is temperature-independent.  Since this entropy must ultimately be removed from the system, the heat dissipated by the cryostat (at room temperature) will be larger by at least $T_h/T_c$: $dQ_h \geq (T_h/T_c) dQ_c$, where $T_h$ is room temperature (in practice the dissipation is much larger due to the low efficiency of most cryostats relative to a Carnot engine \cite{Blotter1993}).  Since this lower limit to $dQ_h$ is independent of $T_c$, reducing Johnson noise by lowering the chip temperature does not confer any benefit in {\it total} energy efficiency.}

\section{Landauer Limit Dependence on Architecture and Bit Precision}
\label{sec:s5}

\begin{table}[b!]
\begin{center}
\begin{tabular}{l|llll|llll}
\hline\hline
                           & \multicolumn{4}{|c|}{Gate Count}   & \multicolumn{4}{c}{Transistor Count} \\ 
                           & 8 bit & 16 bit & 32 bit & 64 bit   & 8 bit & 16 bit & 32 bit & 64 bit \\ \hline
Wallace and Booth          & 33  & 221  &  1077 & 4709          & 168  & 1080   & 5208  & 22680 \\ 
Serial Parallel            & 384 & 1536 & 6144  & 24576         & 1920  & 7680   & 30720  & 122880 \\ 
Braun Multiplier           & 344 & 1456 & 5984  & 24256         & 1728  & 7296   & 29952 & 121344 \\ 
Ripple Carry Adder Based   & 96  & 384  & 1536  & 6144          & 480   & 1920   & 7680  & 30720 \\ 
Vedic Multiplier           & 49  & 281  & 1321  & 5705          & 252   & 1428   & 6660 & 28644 \\ 
\hline\hline
\end{tabular}
\caption{Gate count and transistor count for commonly used integer multipliers.}
\label{tab:ts1}
\end{center}
\end{table}

The Landauer Limit \cite{Landauer1961} sets a thermodynamic lower bound of $kT\log(2)$ for the energy of an irreversible logic gate with a single-bit output.  For a more complex operation such as multiplication, the limit will be $(kT\log(2))\times G$, where $G$ is the gate count.  The estimate of $G = 10^3$ for a MAC, used in the main text, is roughly accurate for 32-bit precision.  However, the trend in deep learning has been toward lower-precision arithmetic, which is faster and more energy-efficient \cite{Jouppi2017}.  Table \ref{tab:ts1} lists the gate counts and transistor counts of commonly used integer multpliers \cite{Nagamatsu1989, Yao1993, Koren2001, Anitha2012, Bewick1994} at all common bit precisions.

\begin{figure}[b!]
\begin{center}
\includegraphics[width=0.55\textwidth]{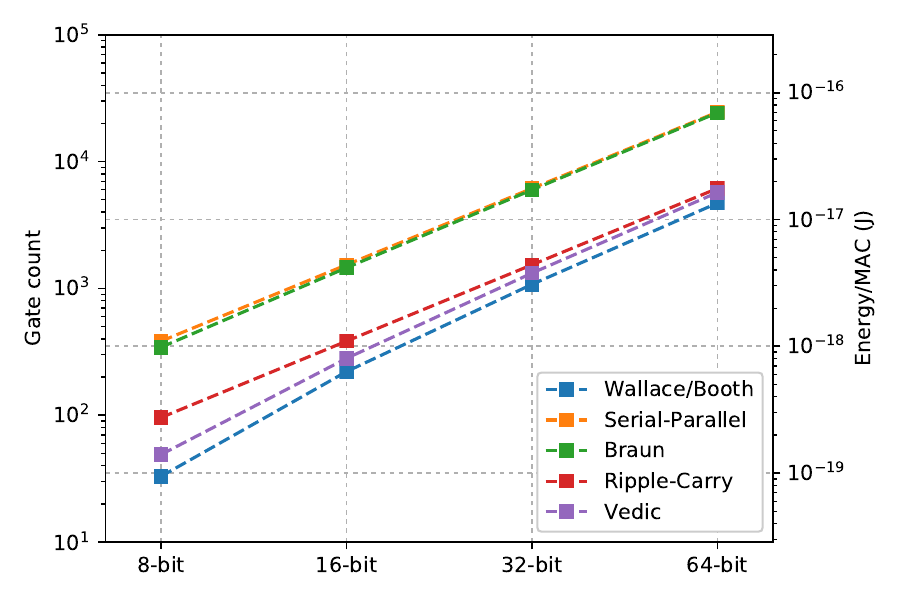}
\caption{Gate count and room-temperature Landauer limit for the multipliers listed in Table \ref{tab:ts1}.}
\label{fig:fs3}
\end{center}
\end{figure}

Gate count is an algorithmic construct independent of the physical implementation.  Any digital computer based on irreversible gates will satisfy the bound $E_{\rm op} \geq (kT\log(2))\times G$, regardless of how (or whether) the gates are implemented with CMOS transistors.  Fig.~\ref{fig:fs3} shows the Landauer limit for the multipliers listed in Table~\ref{tab:ts1}.  Since multipliers require more gates than adders, the multiplier accounts for the majority of the gates for a MAC, and Fig.~\ref{fig:fs3} is a rough lower bound for the Landauer limit to $E_{\rm mac}$.

Note that for all multiplier types the gate and transistor count scale quadratically with number of bits.  By moving from 32-bit arithmetic to 8-bit arithmetic, the Landauer limit decreases by a factor of 16.  For the most efficient multiplier under this metric (Wallace/Booth), it reaches just under 100 zJ/MAC ($10^{-19}$ J/MAC).  This is significantly lower than the 3~aJ value quoted in the main text, but still slightly above the SQL obtained for the larger networks in Fig.~5 (main text), suggesting that even when comparing against low bit-precision digital architectures, the optical network can beat the Landauer limit.

In practice, digital designers use different multipliers depending on if they are optimizing for speed, area on chip, or energy consumption \cite{Bewick1994}.  Wallace Tree based multipliers with Booth encoding have proved to be very energy efficient in modern designs and are common.  Braun Multipliers are sometimes used for unsigned multiplication \cite{Anitha2012}.



\bibliographystyle{naturemag}
\bibliography{PaperRefs}

\begin{thebibliography}{10}
\expandafter\ifx\csname url\endcsname\relax
  \def\url#1{\texttt{#1}}\fi
\expandafter\ifx\csname urlprefix\endcsname\relax\def\urlprefix{URL }\fi
\providecommand{\bibinfo}[2]{#2}
\providecommand{\eprint}[2][]{\url{#2}}

\bibitem{Krizhevsky2012}
\bibinfo{author}{Krizhevsky, A.}, \bibinfo{author}{Sutskever, I.} \&
  \bibinfo{author}{Hinton, G.~E.}
\newblock \bibinfo{title}{Imagenet classification with deep convolutional
  neural networks}.
\newblock In \emph{\bibinfo{booktitle}{Advances in neural information
  processing systems}}, \bibinfo{pages}{1097--1105} (\bibinfo{year}{2012}).

\bibitem{Young2018}
\bibinfo{author}{Young, T.}, \bibinfo{author}{Hazarika, D.},
  \bibinfo{author}{Poria, S.} \& \bibinfo{author}{Cambria, E.}
\newblock \bibinfo{title}{Recent trends in deep learning based natural language
  processing}.
\newblock \emph{\bibinfo{journal}{IEEE Computational Intelligence Magazine}}
  \textbf{\bibinfo{volume}{13}}, \bibinfo{pages}{55--75}
  (\bibinfo{year}{2018}).

\bibitem{Silver2017}
\bibinfo{author}{Silver, D.} \emph{et~al.}
\newblock \bibinfo{title}{Mastering chess and shogi by self-play with a general
  reinforcement learning algorithm}.
\newblock \emph{\bibinfo{journal}{arXiv preprint arXiv:1712.01815}}
  (\bibinfo{year}{2017}).

\bibitem{Gilmer2017}
\bibinfo{author}{Gilmer, J.}, \bibinfo{author}{Schoenholz, S.~S.},
  \bibinfo{author}{Riley, P.~F.}, \bibinfo{author}{Vinyals, O.} \&
  \bibinfo{author}{Dahl, G.~E.}
\newblock \bibinfo{title}{Neural message passing for quantum chemistry}.
\newblock \emph{\bibinfo{journal}{arXiv preprint arXiv:1704.01212}}
  (\bibinfo{year}{2017}).

\bibitem{Wang2016}
\bibinfo{author}{Wang, D.}, \bibinfo{author}{Khosla, A.},
  \bibinfo{author}{Gargeya, R.}, \bibinfo{author}{Irshad, H.} \&
  \bibinfo{author}{Beck, A.~H.}
\newblock \bibinfo{title}{Deep learning for identifying metastatic breast
  cancer}.
\newblock \emph{\bibinfo{journal}{arXiv preprint arXiv:1606.05718}}
  (\bibinfo{year}{2016}).

\bibitem{Rosenblatt1958}
\bibinfo{author}{Rosenblatt, F.}
\newblock \bibinfo{title}{The perceptron: a probabilistic model for information
  storage and organization in the brain.}
\newblock \emph{\bibinfo{journal}{Psychological {R}eview}}
  \textbf{\bibinfo{volume}{65}}, \bibinfo{pages}{386} (\bibinfo{year}{1958}).

\bibitem{Werbos1974}
\bibinfo{author}{Werbos, P.}
\newblock \bibinfo{title}{Beyond regression: New tools for prediction and
  analysis in the behavioral sciences}.
\newblock \emph{\bibinfo{journal}{Ph.D.\ dissertation, Harvard University}}
  (\bibinfo{year}{1974}).

\bibitem{LeCun1998}
\bibinfo{author}{LeCun, Y.}, \bibinfo{author}{Bottou, L.},
  \bibinfo{author}{Bengio, Y.} \& \bibinfo{author}{Haffner, P.}
\newblock \bibinfo{title}{Gradient-based learning applied to document
  recognition}.
\newblock \emph{\bibinfo{journal}{Proceedings of the IEEE}}
  \textbf{\bibinfo{volume}{86}}, \bibinfo{pages}{2278--2324}
  (\bibinfo{year}{1998}).

\bibitem{ILSVRC15}
\bibinfo{author}{Russakovsky, O.} \emph{et~al.}
\newblock \bibinfo{title}{{ImageNet Large Scale Visual Recognition Challenge}}.
\newblock \emph{\bibinfo{journal}{International Journal of Computer Vision
  (IJCV)}} \textbf{\bibinfo{volume}{115}}, \bibinfo{pages}{211--252}
  (\bibinfo{year}{2015}).

\bibitem{Moore1965}
\bibinfo{author}{Moore, G.~E.}
\newblock \bibinfo{title}{Cramming more components onto integrated circuits}.
\newblock \emph{\bibinfo{journal}{Electronics}} \bibinfo{pages}{114--117}
  (\bibinfo{year}{1965}).

\bibitem{Steinkraus2005}
\bibinfo{author}{Steinkraus, D.}, \bibinfo{author}{Buck, I.} \&
  \bibinfo{author}{Simard, P.}
\newblock \bibinfo{title}{Using {GPUs} for machine learning algorithms}.
\newblock In \emph{\bibinfo{booktitle}{Document Analysis and Recognition, 2005.
  Proceedings. Eighth International Conference on}},
  \bibinfo{pages}{1115--1120} (\bibinfo{organization}{IEEE},
  \bibinfo{year}{2005}).

\bibitem{Canziani2017}
\bibinfo{author}{Canziani, A.}, \bibinfo{author}{Culurciello, E.} \&
  \bibinfo{author}{Paszke, A.}
\newblock \bibinfo{title}{Evaluation of neural network architectures for
  embedded systems}.
\newblock In \emph{\bibinfo{booktitle}{Circuits and Systems (ISCAS), 2017 IEEE
  International Symposium on}}, \bibinfo{pages}{1--4}
  (\bibinfo{organization}{IEEE}, \bibinfo{year}{2017}).

\bibitem{Sze2017}
\bibinfo{author}{Sze, V.}, \bibinfo{author}{Chen, Y.-H.},
  \bibinfo{author}{Yang, T.-J.} \& \bibinfo{author}{Emer, J.~S.}
\newblock \bibinfo{title}{Efficient processing of deep neural networks: A
  tutorial and survey}.
\newblock \emph{\bibinfo{journal}{Proceedings of the IEEE}}
  \textbf{\bibinfo{volume}{105}}, \bibinfo{pages}{2295--2329}
  (\bibinfo{year}{2017}).

\bibitem{Horowitz2014}
\bibinfo{author}{Horowitz, M.}
\newblock \bibinfo{title}{Computing's energy problem (and what we can do about
  it)}.
\newblock In \emph{\bibinfo{booktitle}{Solid-State Circuits Conference Digest
  of Technical Papers (ISSCC), 2014 IEEE International}},
  \bibinfo{pages}{10--14} (\bibinfo{organization}{IEEE}, \bibinfo{year}{2014}).

\bibitem{Keckler2011}
\bibinfo{author}{Keckler, S.~W.}, \bibinfo{author}{Dally, W.~J.},
  \bibinfo{author}{Khailany, B.}, \bibinfo{author}{Garland, M.} \&
  \bibinfo{author}{Glasco, D.}
\newblock \bibinfo{title}{{GPUs} and the future of parallel computing}.
\newblock \emph{\bibinfo{journal}{IEEE Micro}} \bibinfo{pages}{7--17}
  (\bibinfo{year}{2011}).

\bibitem{Chen2014}
\bibinfo{author}{Chen, T.} \emph{et~al.}
\newblock \bibinfo{title}{{DianNao}: A small-footprint high-throughput
  accelerator for ubiquitous machine-learning}.
\newblock \emph{\bibinfo{journal}{ACM Sigplan Notices}}
  \textbf{\bibinfo{volume}{49}}, \bibinfo{pages}{269--284}
  (\bibinfo{year}{2014}).

\bibitem{Jouppi2017}
\bibinfo{author}{Jouppi, N.~P.} \emph{et~al.}
\newblock \bibinfo{title}{In-datacenter performance analysis of a tensor
  processing unit}.
\newblock In \emph{\bibinfo{booktitle}{Computer Architecture (ISCA), 2017
  ACM/IEEE 44th Annual International Symposium on}}, \bibinfo{pages}{1--12}
  (\bibinfo{organization}{IEEE}, \bibinfo{year}{2017}).

\bibitem{George2016}
\bibinfo{author}{George, S.} \emph{et~al.}
\newblock \bibinfo{title}{A programmable and configurable mixed-mode {FPAA}
  {SoC}}.
\newblock \emph{\bibinfo{journal}{IEEE Transactions on Very Large Scale
  Integration (VLSI) Systems}} \textbf{\bibinfo{volume}{24}},
  \bibinfo{pages}{2253--2261} (\bibinfo{year}{2016}).

\bibitem{Kim2011}
\bibinfo{author}{Kim, K.-H.} \emph{et~al.}
\newblock \bibinfo{title}{A functional hybrid memristor crossbar-array/{CMOS}
  system for data storage and neuromorphic applications}.
\newblock \emph{\bibinfo{journal}{Nano {L}etters}}
  \textbf{\bibinfo{volume}{12}}, \bibinfo{pages}{389--395}
  (\bibinfo{year}{2011}).

\bibitem{Li2018}
\bibinfo{author}{Li, C.} \emph{et~al.}
\newblock \bibinfo{title}{Efficient and self-adaptive in-situ learning in
  multilayer memristor neural networks}.
\newblock \emph{\bibinfo{journal}{Nature Communications}}
  \textbf{\bibinfo{volume}{9}}, \bibinfo{pages}{2385} (\bibinfo{year}{2018}).

\bibitem{Feinberg2018}
\bibinfo{author}{Feinberg, B.}, \bibinfo{author}{Wang, S.} \&
  \bibinfo{author}{Ipek, E.}
\newblock \bibinfo{title}{Making memristive neural network accelerators
  reliable}.
\newblock In \emph{\bibinfo{booktitle}{2018 IEEE International Symposium on
  High Performance Computer Architecture (HPCA)}}, \bibinfo{pages}{52--65}
  (\bibinfo{organization}{IEEE}, \bibinfo{year}{2018}).

\bibitem{Lin2018}
\bibinfo{author}{Lin, X.} \emph{et~al.}
\newblock \bibinfo{title}{All-optical machine learning using diffractive deep
  neural networks}.
\newblock \emph{\bibinfo{journal}{Science}} \textbf{\bibinfo{volume}{361}},
  \bibinfo{pages}{1004--1008} (\bibinfo{year}{2018}).

\bibitem{Shen2017}
\bibinfo{author}{Shen, Y.} \emph{et~al.}
\newblock \bibinfo{title}{Deep learning with coherent nanophotonic circuits}.
\newblock \emph{\bibinfo{journal}{Nature Photonics}}
  \textbf{\bibinfo{volume}{11}}, \bibinfo{pages}{441} (\bibinfo{year}{2017}).

\bibitem{Tait2017}
\bibinfo{author}{Tait, A.~N.} \emph{et~al.}
\newblock \bibinfo{title}{Neuromorphic photonic networks using silicon photonic
  weight banks}.
\newblock \emph{\bibinfo{journal}{Scientific Reports}}
  \textbf{\bibinfo{volume}{7}}, \bibinfo{pages}{7430} (\bibinfo{year}{2017}).

\bibitem{Kahn2017}
\bibinfo{author}{Kahn, J.~M.} \& \bibinfo{author}{Miller, D.~A.}
\newblock \bibinfo{title}{Communications expands its space}.
\newblock \emph{\bibinfo{journal}{Nature Photonics}}
  \textbf{\bibinfo{volume}{11}}, \bibinfo{pages}{5} (\bibinfo{year}{2017}).

\bibitem{Miller2017}
\bibinfo{author}{Miller, D.~A.}
\newblock \bibinfo{title}{Attojoule optoelectronics for low-energy information
  processing and communications}.
\newblock \emph{\bibinfo{journal}{Journal of Lightwave Technology}}
  \textbf{\bibinfo{volume}{35}}, \bibinfo{pages}{346--396}
  (\bibinfo{year}{2017}).

\bibitem{Rogalski2012}
\bibinfo{author}{Rogalski, A.}
\newblock \bibinfo{title}{Progress in focal plane array technologies}.
\newblock \emph{\bibinfo{journal}{Progress in Quantum Electronics}}
  \textbf{\bibinfo{volume}{36}}, \bibinfo{pages}{342--473}
  (\bibinfo{year}{2012}).

\bibitem{Marandi2014}
\bibinfo{author}{Marandi, A.}, \bibinfo{author}{Wang, Z.},
  \bibinfo{author}{Takata, K.}, \bibinfo{author}{Byer, R.~L.} \&
  \bibinfo{author}{Yamamoto, Y.}
\newblock \bibinfo{title}{Network of time-multiplexed optical parametric
  oscillators as a coherent {I}sing machine}.
\newblock \emph{\bibinfo{journal}{Nature Photonics}}
  \textbf{\bibinfo{volume}{8}}, \bibinfo{pages}{937} (\bibinfo{year}{2014}).

\bibitem{Inagaki2016}
\bibinfo{author}{Inagaki, T.} \emph{et~al.}
\newblock \bibinfo{title}{Large-scale {Ising} spin network based on degenerate
  optical parametric oscillators}.
\newblock \emph{\bibinfo{journal}{Nature Photonics}}
  \textbf{\bibinfo{volume}{10}}, \bibinfo{pages}{415} (\bibinfo{year}{2016}).

\bibitem{McMahon2016}
\bibinfo{author}{McMahon, P.~L.} \emph{et~al.}
\newblock \bibinfo{title}{A fully programmable 100-spin coherent {Ising}
  machine with all-to-all connections}.
\newblock \emph{\bibinfo{journal}{Science}} \textbf{\bibinfo{volume}{354}},
  \bibinfo{pages}{614--617} (\bibinfo{year}{2016}).

\bibitem{Miller2010b}
\bibinfo{author}{Miller, D.~A.}
\newblock \bibinfo{title}{Are optical transistors the logical next step?}
\newblock \emph{\bibinfo{journal}{Nature Photonics}}
  \textbf{\bibinfo{volume}{4}}, \bibinfo{pages}{3} (\bibinfo{year}{2010}).

\bibitem{Tait2014}
\bibinfo{author}{Tait, A.~N.}, \bibinfo{author}{Nahmias, M.~A.},
  \bibinfo{author}{Shastri, B.~J.} \& \bibinfo{author}{Prucnal, P.~R.}
\newblock \bibinfo{title}{Broadcast and weight: an integrated network for
  scalable photonic spike processing}.
\newblock \emph{\bibinfo{journal}{Journal of Lightwave Technology}}
  \textbf{\bibinfo{volume}{32}}, \bibinfo{pages}{3427--3439}
  (\bibinfo{year}{2014}).

\bibitem{Vandoorne2008}
\bibinfo{author}{Vandoorne, K.} \emph{et~al.}
\newblock \bibinfo{title}{Toward optical signal processing using photonic
  reservoir computing}.
\newblock \emph{\bibinfo{journal}{Optics express}}
  \textbf{\bibinfo{volume}{16}}, \bibinfo{pages}{11182--11192}
  (\bibinfo{year}{2008}).

\bibitem{Paquot2012}
\bibinfo{author}{Paquot, Y.} \emph{et~al.}
\newblock \bibinfo{title}{Optoelectronic reservoir computing}.
\newblock \emph{\bibinfo{journal}{Scientific Reports}}
  \textbf{\bibinfo{volume}{2}}, \bibinfo{pages}{287} (\bibinfo{year}{2012}).

\bibitem{Larger2012}
\bibinfo{author}{Larger, L.} \emph{et~al.}
\newblock \bibinfo{title}{Photonic information processing beyond turing: an
  optoelectronic implementation of reservoir computing}.
\newblock \emph{\bibinfo{journal}{Optics express}}
  \textbf{\bibinfo{volume}{20}}, \bibinfo{pages}{3241--3249}
  (\bibinfo{year}{2012}).

\bibitem{Nahmias2013}
\bibinfo{author}{Nahmias, M.~A.}, \bibinfo{author}{Shastri, B.~J.},
  \bibinfo{author}{Tait, A.~N.} \& \bibinfo{author}{Prucnal, P.~R.}
\newblock \bibinfo{title}{A leaky integrate-and-fire laser neuron for ultrafast
  cognitive computing}.
\newblock \emph{\bibinfo{journal}{IEEE journal of selected topics in quantum
  electronics}} \textbf{\bibinfo{volume}{19}}, \bibinfo{pages}{1--12}
  (\bibinfo{year}{2013}).

\bibitem{Brunner2016}
\bibinfo{author}{Brunner, D.}, \bibinfo{author}{Reitzenstein, S.} \&
  \bibinfo{author}{Fischer, I.}
\newblock \bibinfo{title}{All-optical neuromorphic computing in optical
  networks of semiconductor lasers}.
\newblock In \emph{\bibinfo{booktitle}{2016 IEEE International Conference on
  Rebooting Computing (ICRC)}}, \bibinfo{pages}{1--2}
  (\bibinfo{organization}{IEEE}, \bibinfo{year}{2016}).

\bibitem{Reck1994}
\bibinfo{author}{Reck, M.}, \bibinfo{author}{Zeilinger, A.},
  \bibinfo{author}{Bernstein, H.~J.} \& \bibinfo{author}{Bertani, P.}
\newblock \bibinfo{title}{Experimental realization of any discrete unitary
  operator}.
\newblock \emph{\bibinfo{journal}{Physical Review Letters}}
  \textbf{\bibinfo{volume}{73}}, \bibinfo{pages}{58} (\bibinfo{year}{1994}).

\bibitem{Clements2016}
\bibinfo{author}{Clements, W.~R.}, \bibinfo{author}{Humphreys, P.~C.},
  \bibinfo{author}{Metcalf, B.~J.}, \bibinfo{author}{Kolthammer, W.~S.} \&
  \bibinfo{author}{Walmsley, I.~A.}
\newblock \bibinfo{title}{Optimal design for universal multiport
  interferometers}.
\newblock \emph{\bibinfo{journal}{Optica}} \textbf{\bibinfo{volume}{3}},
  \bibinfo{pages}{1460--1465} (\bibinfo{year}{2016}).

\bibitem{Einstein1905}
\bibinfo{author}{Einstein, A.}
\newblock \bibinfo{title}{{\"U}ber einen die erzeugung und verwandlung des
  lichtes betreffenden heuristischen gesichtspunkt}.
\newblock \emph{\bibinfo{journal}{Annalen der Physik}}
  \textbf{\bibinfo{volume}{322}}, \bibinfo{pages}{132--148}
  (\bibinfo{year}{1905}).

\bibitem{WallsMilburn}
\bibinfo{author}{Walls, D.~F.} \& \bibinfo{author}{Milburn, G.~J.}
\newblock \emph{\bibinfo{title}{Quantum {O}ptics}}
  (\bibinfo{publisher}{Springer Science \& Business Media},
  \bibinfo{year}{2007}).

\bibitem{Hayat1996}
\bibinfo{author}{Hayat, M.~M.}, \bibinfo{author}{Saleh, B.~E.} \&
  \bibinfo{author}{Gubner, J.~A.}
\newblock \bibinfo{title}{Shot-noise-limited performance of optical neural
  networks}.
\newblock \emph{\bibinfo{journal}{IEEE Transactions on Neural Networks}}
  \textbf{\bibinfo{volume}{7}}, \bibinfo{pages}{700--708}
  (\bibinfo{year}{1996}).

\bibitem{Caves1981}
\bibinfo{author}{Caves, C.~M.}
\newblock \bibinfo{title}{Quantum-mechanical noise in an interferometer}.
\newblock \emph{\bibinfo{journal}{Physical Review D}}
  \textbf{\bibinfo{volume}{23}}, \bibinfo{pages}{1693} (\bibinfo{year}{1981}).

\bibitem{Jaekel1990}
\bibinfo{author}{Jaekel, M.~T.} \& \bibinfo{author}{Reynaud, S.}
\newblock \bibinfo{title}{Quantum limits in interferometric measurements}.
\newblock \emph{\bibinfo{journal}{EPL (Europhysics Letters)}}
  \textbf{\bibinfo{volume}{13}}, \bibinfo{pages}{301} (\bibinfo{year}{1990}).

\bibitem{Grote2013}
\bibinfo{author}{Grote, H.} \emph{et~al.}
\newblock \bibinfo{title}{First long-term application of squeezed states of
  light in a gravitational-wave observatory}.
\newblock \emph{\bibinfo{journal}{Physical Review Letters}}
  \textbf{\bibinfo{volume}{110}}, \bibinfo{pages}{181101}
  (\bibinfo{year}{2013}).

\bibitem{Holmstrom1992}
\bibinfo{author}{Holmstrom, L.} \& \bibinfo{author}{Koistinen, P.}
\newblock \bibinfo{title}{Using additive noise in back-propagation training}.
\newblock \emph{\bibinfo{journal}{IEEE transactions on neural networks}}
  \textbf{\bibinfo{volume}{3}}, \bibinfo{pages}{24--38} (\bibinfo{year}{1992}).

\bibitem{Hinton2012}
\bibinfo{author}{Hinton, G.~E.}, \bibinfo{author}{Srivastava, N.},
  \bibinfo{author}{Krizhevsky, A.}, \bibinfo{author}{Sutskever, I.} \&
  \bibinfo{author}{Salakhutdinov, R.~R.}
\newblock \bibinfo{title}{Improving neural networks by preventing co-adaptation
  of feature detectors}.
\newblock \emph{\bibinfo{journal}{arXiv preprint arXiv:1207.0580}}
  (\bibinfo{year}{2012}).

\bibitem{Gibbs2012}
\bibinfo{author}{Gibbs, H.}
\newblock \emph{\bibinfo{title}{Optical bistability: controlling light with
  light}} (\bibinfo{publisher}{Elsevier}, \bibinfo{year}{2012}).

\bibitem{Savage1988}
\bibinfo{author}{Savage, C.} \& \bibinfo{author}{Carmichael, H.}
\newblock \bibinfo{title}{Single atom optical bistability}.
\newblock \emph{\bibinfo{journal}{IEEE Journal of Quantum Electronics}}
  \textbf{\bibinfo{volume}{24}}, \bibinfo{pages}{1495--1498}
  (\bibinfo{year}{1988}).

\bibitem{Santori2014}
\bibinfo{author}{Santori, C.} \emph{et~al.}
\newblock \bibinfo{title}{Quantum noise in large-scale coherent nonlinear
  photonic circuits}.
\newblock \emph{\bibinfo{journal}{Physical Review Applied}}
  \textbf{\bibinfo{volume}{1}}, \bibinfo{pages}{054005} (\bibinfo{year}{2014}).

\bibitem{Kerckhoff2011}
\bibinfo{author}{Kerckhoff, J.}, \bibinfo{author}{Armen, M.~A.} \&
  \bibinfo{author}{Mabuchi, H.}
\newblock \bibinfo{title}{Remnants of semiclassical bistability in the
  few-photon regime of cavity {QED}}.
\newblock \emph{\bibinfo{journal}{Optics Express}}
  \textbf{\bibinfo{volume}{19}}, \bibinfo{pages}{24468--24482}
  (\bibinfo{year}{2011}).

\bibitem{Ji2017}
\bibinfo{author}{Ji, X.} \emph{et~al.}
\newblock \bibinfo{title}{Ultra-low-loss on-chip resonators with sub-milliwatt
  parametric oscillation threshold}.
\newblock \emph{\bibinfo{journal}{Optica}} \textbf{\bibinfo{volume}{4}},
  \bibinfo{pages}{619--624} (\bibinfo{year}{2017}).

\bibitem{Hu2018}
\bibinfo{author}{Hu, S.} \emph{et~al.}
\newblock \bibinfo{title}{Experimental realization of deep-subwavelength
  confinement in dielectric optical resonators}.
\newblock \emph{\bibinfo{journal}{Science Advances}}
  \textbf{\bibinfo{volume}{4}}, \bibinfo{pages}{eaat2355}
  (\bibinfo{year}{2018}).

\bibitem{Wang2018}
\bibinfo{author}{Wang, C.} \emph{et~al.}
\newblock \bibinfo{title}{Ultrahigh-efficiency second-harmonic generation in
  nanophotonic {PPLN} waveguides}.
\newblock \emph{\bibinfo{journal}{arXiv preprint arXiv:1810.09235}}
  (\bibinfo{year}{2018}).

\bibitem{Miller2012}
\bibinfo{author}{Miller, D.~A.}
\newblock \bibinfo{title}{Energy consumption in optical modulators for
  interconnects}.
\newblock \emph{\bibinfo{journal}{Optics Express}}
  \textbf{\bibinfo{volume}{20}}, \bibinfo{pages}{A293--A308}
  (\bibinfo{year}{2012}).

\bibitem{Sun2015}
\bibinfo{author}{Sun, C.} \emph{et~al.}
\newblock \bibinfo{title}{Single-chip microprocessor that communicates directly
  using light}.
\newblock \emph{\bibinfo{journal}{Nature}} \textbf{\bibinfo{volume}{528}},
  \bibinfo{pages}{534} (\bibinfo{year}{2015}).

\bibitem{Atabaki2018}
\bibinfo{author}{Atabaki, A.~H.} \emph{et~al.}
\newblock \bibinfo{title}{Integrating photonics with silicon nanoelectronics
  for the next generation of systems on a chip}.
\newblock \emph{\bibinfo{journal}{Nature}} \textbf{\bibinfo{volume}{556}},
  \bibinfo{pages}{349} (\bibinfo{year}{2018}).

\bibitem{Michaels2018}
\bibinfo{author}{Michaels, A.} \& \bibinfo{author}{Yablonovitch, E.}
\newblock \bibinfo{title}{Inverse design of near unity efficiency perfectly
  vertical grating couplers}.
\newblock \emph{\bibinfo{journal}{Optics Express}}
  \textbf{\bibinfo{volume}{26}}, \bibinfo{pages}{4766--4779}
  (\bibinfo{year}{2018}).

\bibitem{Saeedi2016}
\bibinfo{author}{Saeedi, S.}, \bibinfo{author}{Menezo, S.},
  \bibinfo{author}{Pares, G.} \& \bibinfo{author}{Emami, A.}
\newblock \bibinfo{title}{A 25 {Gb}/s {3D}-integrated {CMOS}/silicon-photonic
  receiver for low-power high-sensitivity optical communication}.
\newblock \emph{\bibinfo{journal}{Journal of Lightwave Technology}}
  \textbf{\bibinfo{volume}{34}}, \bibinfo{pages}{2924--2933}
  (\bibinfo{year}{2016}).

\bibitem{Jonsson2011}
\bibinfo{author}{Jonsson, B.~E.}
\newblock \bibinfo{title}{An empirical approach to finding energy efficient
  {ADC} architectures}.
\newblock In \emph{\bibinfo{booktitle}{Proc. of 2011 IMEKO IWADC \& IEEE ADC
  Forum}}, \bibinfo{pages}{1--6} (\bibinfo{year}{2011}).

\bibitem{Notomi2014}
\bibinfo{author}{Notomi, M.}, \bibinfo{author}{Nozaki, K.},
  \bibinfo{author}{Shinya, A.}, \bibinfo{author}{Matsuo, S.} \&
  \bibinfo{author}{Kuramochi, E.}
\newblock \bibinfo{title}{Toward {fJ}/bit optical communication in a chip}.
\newblock \emph{\bibinfo{journal}{Optics Communications}}
  \textbf{\bibinfo{volume}{314}}, \bibinfo{pages}{3--17}
  (\bibinfo{year}{2014}).

\bibitem{Timurdogan2014}
\bibinfo{author}{Timurdogan, E.} \emph{et~al.}
\newblock \bibinfo{title}{An ultralow power athermal silicon modulator}.
\newblock \emph{\bibinfo{journal}{Nature Communications}}
  \textbf{\bibinfo{volume}{5}}, \bibinfo{pages}{4008} (\bibinfo{year}{2014}).

\bibitem{Koos2016}
\bibinfo{author}{Koos, C.} \emph{et~al.}
\newblock \bibinfo{title}{Silicon-organic hybrid ({SOH}) and plasmonic-organic
  hybrid ({POH}) integration}.
\newblock \emph{\bibinfo{journal}{Journal of Lightwave Technology}}
  \textbf{\bibinfo{volume}{34}}, \bibinfo{pages}{256--268}
  (\bibinfo{year}{2016}).

\bibitem{Haffner2018}
\bibinfo{author}{Haffner, C.} \emph{et~al.}
\newblock \bibinfo{title}{Low-loss plasmon-assisted electro-optic modulator}.
\newblock \emph{\bibinfo{journal}{Nature}} \textbf{\bibinfo{volume}{556}},
  \bibinfo{pages}{483} (\bibinfo{year}{2018}).

\bibitem{Srinivasan2016}
\bibinfo{author}{Srinivasan, S.~A.} \emph{et~al.}
\newblock \bibinfo{title}{56 {Gb/s} germanium waveguide electro-absorption
  modulator}.
\newblock \emph{\bibinfo{journal}{Journal of Lightwave Technology}}
  \textbf{\bibinfo{volume}{34}}, \bibinfo{pages}{419--424}
  (\bibinfo{year}{2016}).

\bibitem{Nozaki2016}
\bibinfo{author}{Nozaki, K.} \emph{et~al.}
\newblock \bibinfo{title}{Photonic-crystal nano-photodetector with ultrasmall
  capacitance for on-chip light-to-voltage conversion without an amplifier}.
\newblock \emph{\bibinfo{journal}{Optica}} \textbf{\bibinfo{volume}{3}},
  \bibinfo{pages}{483--492} (\bibinfo{year}{2016}).

\bibitem{Ishi2005}
\bibinfo{author}{Ishi, T.}, \bibinfo{author}{Fujikata, J.},
  \bibinfo{author}{Makita, K.}, \bibinfo{author}{Baba, T.} \&
  \bibinfo{author}{Ohashi, K.}
\newblock \bibinfo{title}{Si nano-photodiode with a surface plasmon antenna}.
\newblock \emph{\bibinfo{journal}{Japanese Journal of Applied Physics}}
  \textbf{\bibinfo{volume}{44}}, \bibinfo{pages}{L364} (\bibinfo{year}{2005}).

\bibitem{Tang2008}
\bibinfo{author}{Tang, L.} \emph{et~al.}
\newblock \bibinfo{title}{Nanometre-scale germanium photodetector enhanced by a
  near-infrared dipole antenna}.
\newblock \emph{\bibinfo{journal}{Nature Photonics}}
  \textbf{\bibinfo{volume}{2}}, \bibinfo{pages}{226} (\bibinfo{year}{2008}).

\bibitem{Cao2010}
\bibinfo{author}{Cao, L.}, \bibinfo{author}{Park, J.-S.}, \bibinfo{author}{Fan,
  P.}, \bibinfo{author}{Clemens, B.} \& \bibinfo{author}{Brongersma, M.~L.}
\newblock \bibinfo{title}{Resonant germanium nanoantenna photodetectors}.
\newblock \emph{\bibinfo{journal}{Nano letters}} \textbf{\bibinfo{volume}{10}},
  \bibinfo{pages}{1229--1233} (\bibinfo{year}{2010}).

\bibitem{Pierce1956}
\bibinfo{author}{Pierce, J.}
\newblock \bibinfo{title}{Physical sources of noise}.
\newblock \emph{\bibinfo{journal}{Proceedings of the IRE}}
  \textbf{\bibinfo{volume}{44}}, \bibinfo{pages}{601--608}
  (\bibinfo{year}{1956}).

\bibitem{Landauer1961}
\bibinfo{author}{Landauer, R.}
\newblock \bibinfo{title}{Irreversibility and heat generation in the computing
  process}.
\newblock \emph{\bibinfo{journal}{IBM Journal of Research and Development}}
  \textbf{\bibinfo{volume}{5}}, \bibinfo{pages}{183--191}
  (\bibinfo{year}{1961}).

\bibitem{Nagamatsu1989}
\bibinfo{author}{Nagamatsu, M.}, \bibinfo{author}{Tanaka, S.},
  \bibinfo{author}{Mori, J.}, \bibinfo{author}{Noguchi, T.} \&
  \bibinfo{author}{Hatanaka, K.}
\newblock \bibinfo{title}{A 15-ns 32$\times$32-bit {CMOS} multiplier with an
  improved parallel structure}.
\newblock In \emph{\bibinfo{booktitle}{Custom Integrated Circuits Conference,
  1989., Proceedings of the IEEE 1989}}, \bibinfo{pages}{10--3}
  (\bibinfo{organization}{IEEE}, \bibinfo{year}{1989}).

\bibitem{Yao1993}
\bibinfo{author}{Yao, H.~H.} \& \bibinfo{author}{Swartzlander, E.}
\newblock \bibinfo{title}{Serial-parallel multipliers}.
\newblock In \emph{\bibinfo{booktitle}{Signals, Systems and Computers, 1993.
  1993 Conference Record of The Twenty-Seventh Asilomar Conference on}},
  \bibinfo{pages}{359--363} (\bibinfo{organization}{IEEE},
  \bibinfo{year}{1993}).

\bibitem{Lawson1979}
\bibinfo{author}{Lawson, C.~L.}, \bibinfo{author}{Hanson, R.~J.},
  \bibinfo{author}{Kincaid, D.~R.} \& \bibinfo{author}{Krogh, F.~T.}
\newblock \bibinfo{title}{Basic linear algebra subprograms for {Fortran}
  usage}.
\newblock \emph{\bibinfo{journal}{ACM Transactions on Mathematical Software
  (TOMS)}} \textbf{\bibinfo{volume}{5}}, \bibinfo{pages}{308--323}
  (\bibinfo{year}{1979}).

\bibitem{Chetlur2014}
\bibinfo{author}{Chetlur, S.} \emph{et~al.}
\newblock \bibinfo{title}{{cuDNN}: Efficient primitives for deep learning}.
\newblock \emph{\bibinfo{journal}{arXiv preprint arXiv:1410.0759}}
  (\bibinfo{year}{2014}).

\bibitem{Li2016}
\bibinfo{author}{Li, X.}, \bibinfo{author}{Zhang, G.}, \bibinfo{author}{Huang,
  H.~H.}, \bibinfo{author}{Wang, Z.} \& \bibinfo{author}{Zheng, W.}
\newblock \bibinfo{title}{Performance analysis of {GPU}-based convolutional
  neural networks}.
\newblock In \emph{\bibinfo{booktitle}{Parallel Processing (ICPP), 2016 45th
  International Conference on}}, \bibinfo{pages}{67--76}
  (\bibinfo{organization}{IEEE}, \bibinfo{year}{2016}).

\bibitem{Chen2017}
\bibinfo{author}{Chen, Y.-H.}, \bibinfo{author}{Krishna, T.},
  \bibinfo{author}{Emer, J.~S.} \& \bibinfo{author}{Sze, V.}
\newblock \bibinfo{title}{Eyeriss: An energy-efficient reconfigurable
  accelerator for deep convolutional neural networks}.
\newblock \emph{\bibinfo{journal}{IEEE Journal of Solid-State Circuits}}
  \textbf{\bibinfo{volume}{52}}, \bibinfo{pages}{127--138}
  (\bibinfo{year}{2017}).

\bibitem{Bagherian2018}
\bibinfo{author}{Bagherian, H.} \emph{et~al.}
\newblock \bibinfo{title}{On-chip optical convolutional neural networks}.
\newblock \emph{\bibinfo{journal}{arXiv preprint arXiv:1808.03303}}
  (\bibinfo{year}{2018}).

\bibitem{Lugt1964}
\bibinfo{author}{Lugt, A.~V.}
\newblock \bibinfo{title}{Signal detection by complex spatial filtering}.
\newblock \emph{\bibinfo{journal}{IEEE Transactions on information theory}}
  \textbf{\bibinfo{volume}{10}}, \bibinfo{pages}{139--145}
  (\bibinfo{year}{1964}).

\bibitem{Paek1987}
\bibinfo{author}{Paek, E.~G.} \& \bibinfo{author}{Psaltis, D.}
\newblock \bibinfo{title}{Optical associative memory using {Fourier} transform
  holograms}.
\newblock \emph{\bibinfo{journal}{Optical Engineering}}
  \textbf{\bibinfo{volume}{26}}, \bibinfo{pages}{265428}
  (\bibinfo{year}{1987}).

\bibitem{New2017}
\bibinfo{author}{New, N.~J.}
\newblock \bibinfo{title}{Reconfigurable optical processing system}
  (\bibinfo{year}{2017}).
\newblock \bibinfo{note}{{US} {P}atent 9,594,394}.

\bibitem{Chang2018}
\bibinfo{author}{Chang, J.}, \bibinfo{author}{Sitzmann, V.},
  \bibinfo{author}{Dun, X.}, \bibinfo{author}{Heidrich, W.} \&
  \bibinfo{author}{Wetzstein, G.}
\newblock \bibinfo{title}{Hybrid optical-electronic convolutional neural
  networks with optimized diffractive optics for image classification}.
\newblock \emph{\bibinfo{journal}{Scientific Reports}}
  \textbf{\bibinfo{volume}{8}}, \bibinfo{pages}{12324} (\bibinfo{year}{2018}).

\bibitem{Sun2013}
\bibinfo{author}{Sun, J.}, \bibinfo{author}{Timurdogan, E.},
  \bibinfo{author}{Yaacobi, A.}, \bibinfo{author}{Hosseini, E.~S.} \&
  \bibinfo{author}{Watts, M.~R.}
\newblock \bibinfo{title}{Large-scale nanophotonic phased array}.
\newblock \emph{\bibinfo{journal}{Nature}} \textbf{\bibinfo{volume}{493}},
  \bibinfo{pages}{195} (\bibinfo{year}{2013}).

\bibitem{Chung2018}
\bibinfo{author}{Chung, S.}, \bibinfo{author}{Abediasl, H.} \&
  \bibinfo{author}{Hashemi, H.}
\newblock \bibinfo{title}{A monolithically integrated large-scale optical
  phased array in silicon-on-insulator {CMOS}}.
\newblock \emph{\bibinfo{journal}{IEEE Journal of Solid-State Circuits}}
  \textbf{\bibinfo{volume}{53}}, \bibinfo{pages}{275--296}
  (\bibinfo{year}{2018}).

\bibitem{Phare2018}
\bibinfo{author}{Phare, C.~T.}, \bibinfo{author}{Shin, M.~C.},
  \bibinfo{author}{Miller, S.~A.}, \bibinfo{author}{Stern, B.} \&
  \bibinfo{author}{Lipson, M.}
\newblock \bibinfo{title}{Silicon optical phased array with high-efficiency
  beam formation over 180 degree field of view}.
\newblock \emph{\bibinfo{journal}{arXiv preprint arXiv:1802.04624}}
  (\bibinfo{year}{2018}).

\bibitem{Smith1966}
\bibinfo{author}{Smith, W.~J.}
\newblock \emph{\bibinfo{title}{Modern optical engineering}}
  (\bibinfo{publisher}{Tata McGraw-Hill Education}, \bibinfo{year}{1966}).

\bibitem{Peng2018}
\bibinfo{author}{Peng, Y.}
\newblock \bibinfo{title}{Implementation of {AlexNet} with {Tensorflow}}.
\newblock
  \bibinfo{howpublished}{\url{https://github.com/ykpengba/AlexNet-A-Practical-Implementation}}
  (\bibinfo{year}{2018}).

\bibitem{Blotter1993}
\bibinfo{author}{Blotter, P.} \& \bibinfo{author}{Batty, J.}
\newblock \bibinfo{title}{Thermal and mechanical design of cryogenic cooling
  systems}.
\newblock \emph{\bibinfo{journal}{The Infrared and Electro-Optical Systems
  Handbook}} \textbf{\bibinfo{volume}{3}}, \bibinfo{pages}{343--433}
  (\bibinfo{year}{1993}).

\bibitem{Koren2001}
\bibinfo{author}{Koren, I.}
\newblock \emph{\bibinfo{title}{Computer arithmetic algorithms}}
  (\bibinfo{publisher}{AK Peters/CRC Press}, \bibinfo{year}{2001}).

\bibitem{Anitha2012}
\bibinfo{author}{Anitha, R.}, \bibinfo{author}{Nelapati, A.},
  \bibinfo{author}{Lincy~Jesima, W.} \& \bibinfo{author}{Bagyaveereswaran, V.}
\newblock \bibinfo{title}{Comparative study of high performance {B}raun's
  multiplier using {FPGA}}.
\newblock \emph{\bibinfo{journal}{IOSR J Electron Commun Eng (IOSRJECE)}}
  \textbf{\bibinfo{volume}{1}}, \bibinfo{pages}{33--37} (\bibinfo{year}{2012}).

\bibitem{Bewick1994}
\bibinfo{author}{Bewick, G.~W.}
\newblock \emph{\bibinfo{title}{Fast multiplication: algorithms and
  implementation}}.
\newblock Ph.D. thesis, \bibinfo{school}{Stanford University}
  (\bibinfo{year}{1994}).

\end{thebibliography}

\comment{
\begin{table}
\centering
\caption{This is a table with scientific results.}
\medskip
\begin{tabular}{ccccc}
\hline
1 & 2 & 3 & 4 & 5\\
\hline
aaa & bbb & ccc & ddd & eee\\
aaaa & bbbb & cccc & dddd & eeee\\
aaaaa & bbbbb & ccccc & ddddd & eeeee\\
aaaaaa & bbbbbb & cccccc & dddddd & eeeeee\\
1.000 & 2.000 & 3.000 & 4.000 & 5.000\\
\hline
\end{tabular}
\end{table}
}

\end{document}